\documentclass{rspublic}

\usepackage{graphicx}

\usepackage{xcolor}

\definecolor{grey}{gray}{0.5}
\definecolor{fuchsia}{rgb}{255,0,255}

\DeclareSymbolFont{extraup}{U}{zavm}{m}{n}
\DeclareMathSymbol{\vardiamond}{\mathalpha}{extraup}{87}

\DeclareSymbolFont{extraup}{U}{zavm}{m}{n}
\DeclareMathSymbol{\vardiamond}{\mathalpha}{extraup}{87}

\usepackage{wasysym}

\usepackage{amsmath}

\newcommand{\be}{\begin{equation}}
\newcommand{\ee}{\end{equation}}
\newcommand{\bea}{\begin{eqnarray}}
\newcommand{\eea}{\end{eqnarray}}

\newcommand{\ba}{\begin{array}}
\newcommand{\ea}{\end{array}}

\newcommand{\xD}{{\bf D}}

\newcommand{\hx}{\hat{x}}

\newcommand{\htt}{\hat{t}}
\newcommand{\ex}{{\bf e}_x}

\newcommand{\xo}{\mathbf{\omega}}

\newcommand{\xu}{{\bf u}}
\newcommand{\xq}{{\bf q}}

\newcommand{\xn}{{\bf n}}

\newcommand{\etal}{{\it et al.}~}

\newcommand{\Pe}{\mbox{Pe}}
\newcommand{\Pem}{(2/3)\mbox{Pe}}

\def\Var{\mathop{\rm Var}}

\def\om{{\mathbf{\omega}}}

\def\p{{\mathbf{p}}}

\usepackage[square]{natbib}

\begin{document}

\title[Dispersion of algae in channel flow]{Dispersion of swimming algae in laminar and turbulent channel flows: consequences for photobioreactors}

\author[O. A. Croze \etal]{Ottavio A. Croze$^{1*}$, Gaetano Sardina$^{2,3}$, Mansoor Ahmed$^2$, Martin A. Bees$^4$ and Luca Brandt$^2$}
\affiliation{$^1$Department of Plant Sciences, University of Cambridge\\  Cambridge CB2 3EA, UK\\
$^2$Linn\`{e} Flow Centre, SeRC, KTH Mechanics, SE-100 44 Stockholm, Sweden\\
$^3$Facolt\`{a} di Ingegneria, Architettura e Scienze Motorie, Unviversit\`{a} degli Studi di Enna ``Kore'', 94100 Enna, Italy\\
$^4$Department of Mathematics, University of York, York YO10 5DD, UK\\
$^*$Corresponding author: o.croze@physics.org}

\date{}

\label{firstpage}

\maketitle


\begin{abstract}{algae, swimming microorganisms, Taylor dispersion, DNS, turbulent transport, bioreactors}Shear flow significantly affects the transport of swimming algae in suspension. For example, viscous and gravitational torques bias bottom-heavy cells to swim towards regions of downwelling fluid (gyrotaxis). It is necessary to understand how such biases affect algal dispersion in natural and industrial flows, especially in view of growing interest in algal photobioreactors. Motivated by this, we here study the dispersion of gyrotactic algae in laminar and turbulent channel flows using direct numerical simulation (DNS) and the analytical swimming dispersion theory of Bees \& Croze \citep{BeesCroze10}. Time-resolved dispersion measures are evaluated as functions of the P\'{e}clet and Reynolds numbers in upwelling and downwelling flows.  For laminar flows, DNS results are compared with theory using competing descriptions of biased swimming cells in shear flow.  Excellent agreement is found for predictions that employ generalized-Taylor-dispersion.  The results highlight peculiarities of gyrotactic swimmer dispersion relative to passive tracers. In laminar downwelling flow the cell distribution drifts in excess of the mean flow, increasing in magnitude with Peclet number. The cell effective axial diffusivity increases and decreases with Peclet number (for tracers it merely increases). In turbulent flows, gyrotactic effects are weaker, but discernable and manifested as non-zero drift.  These results should significantly impact photobioreactor design.
\end{abstract}


\section{Introduction \label{Intro}}

In natural bodies of water and in industrial bioreactors, microscopic algae experience laminar and turbulent flows that play a critical role in their dispersion, proliferation and productivity \citep{Round83, Greenwelletal10}. The dispersion of passive tracers in fluid flows is well understood, particularly in the industrially relevant cases of flows in pipes and channels \citep{BrennerEdwards93}.  G. I. Taylor first realised that the dispersion of passive tracers in a pipe, caused by the combination of fluid shear and molecular diffusion, could be described in terms of an effective axial diffusivity \citep{Taylor53}. A similar result also holds for turbulent pipe \citep{Taylor54a} and channel \citep{Elder59} flows. Taylor's pioneering analyses of dispersion in a pipe inspired a series of studies that placed the understanding of the dispersion of neutrally buoyant tracers on a firm footing \citep{Aris56, Fisher73}. Particles whose density differs from the suspending medium exhibit more complex behaviour (for example, they can accumulate at walls in turbulent channel flows; see \citep{Picanoetal09, Sardinaetal12} and references therein).

Swimming single-celled algae are known to respond non-trivially to flows. For example, the mean swimming direction of biflagellates, such as {\it Chlamydomonas} and {\it Dunaliella} spp., is biased by imposed flows \citep{Kessler85, PedleyKessler92}. This bias, known as gyrotaxis, results from the combination of viscous torques on the cell body, due to flow gradients, and gravitational torques, arising from bottom-heaviness and sedimentation \citep{OMalleyBees11}. In the absence of flow gradients, the gravitational torque leads cells to swim upwards on average (gravitaxis). Recent simulations and laboratory experiments have shown how inertia and gyrotaxis can play significant roles in the formation of patchiness and/or thin layers of toxic algae in the ocean \citep{Reigadaetal02,Durhametal09, HoeckerMartinezSmyth12}. Gravitaxis and gyrotaxis can also couple in a complex fashion to other biases due to chemical gradients (chemotaxis) and light (phototaxis \citep{WilliamsBees11}).  Therefore, the fact that cells can actively swim across streamlines and accumulate in specific regions of a flow forewarns that the resultant dispersion of biased swimming algae in a complex flow field is nontrivial.

In a recent study, Bees \& Croze extend the classical Taylor-Aris analysis of dispersion in a laminar flow in a tube to the case of biased swimming algae \citep{BeesCroze10}. The Bees \& Croze dispersion theory provides a general theoretical framework to describe the dispersion of biased micro-swimmers in confined geometries, such as pipes and channels. To make predictions for the dispersion of particular organisms, specific functional expressions for the swimming characteristics need to be obtained from microscopic models. For example, using expressions for the mean swimming orientation ${\bf q}$ and diffusivity ${\bf D}$ from the so-called Fokker-Planck model \citep{PedleyKessler92}, Bees \& Croze found that gyrotaxis can significantly modify the axial dispersion of swimming algae (such as {\it Chlamydomonas} spp.) in a pipe flow relative to the case of passive tracers (or dissolved chemicals) \citep{BeesCroze10}. More recently, Bearon {\it et al.} \citep{BearonBeesCroze12} obtained predictions for swimming dispersion in laminar flows in tubes of circular cross-section by using an alternative microscopic model known as generalised Taylor dispersion (GTD), formulated for swimming algae \citep{HillBees02, ManelaFrankel03} and bacteria \citep{Bearon03}. The GTD model is considered superior, as it incorporates correlations in cell positions due to cell locomotion within local a flow, as well as the gyrotactic orientational biases considered in the Fokker-Planck description. The qualitatively distinct predictions, however, have yet to be tested.  Direct tests of the models might involve microscopic tracking of isolated cells swimming in prescribed flows. The Bees \& Croze theory allows the predictions of the two models to be tested for entire populations in macroscopic experiments \citep{CrozeBearonBees12} and individual-based numerical simulations, as we shall demonstrate in this paper.

\begin{figure}[h!]
\centering
\includegraphics[width=84mm]{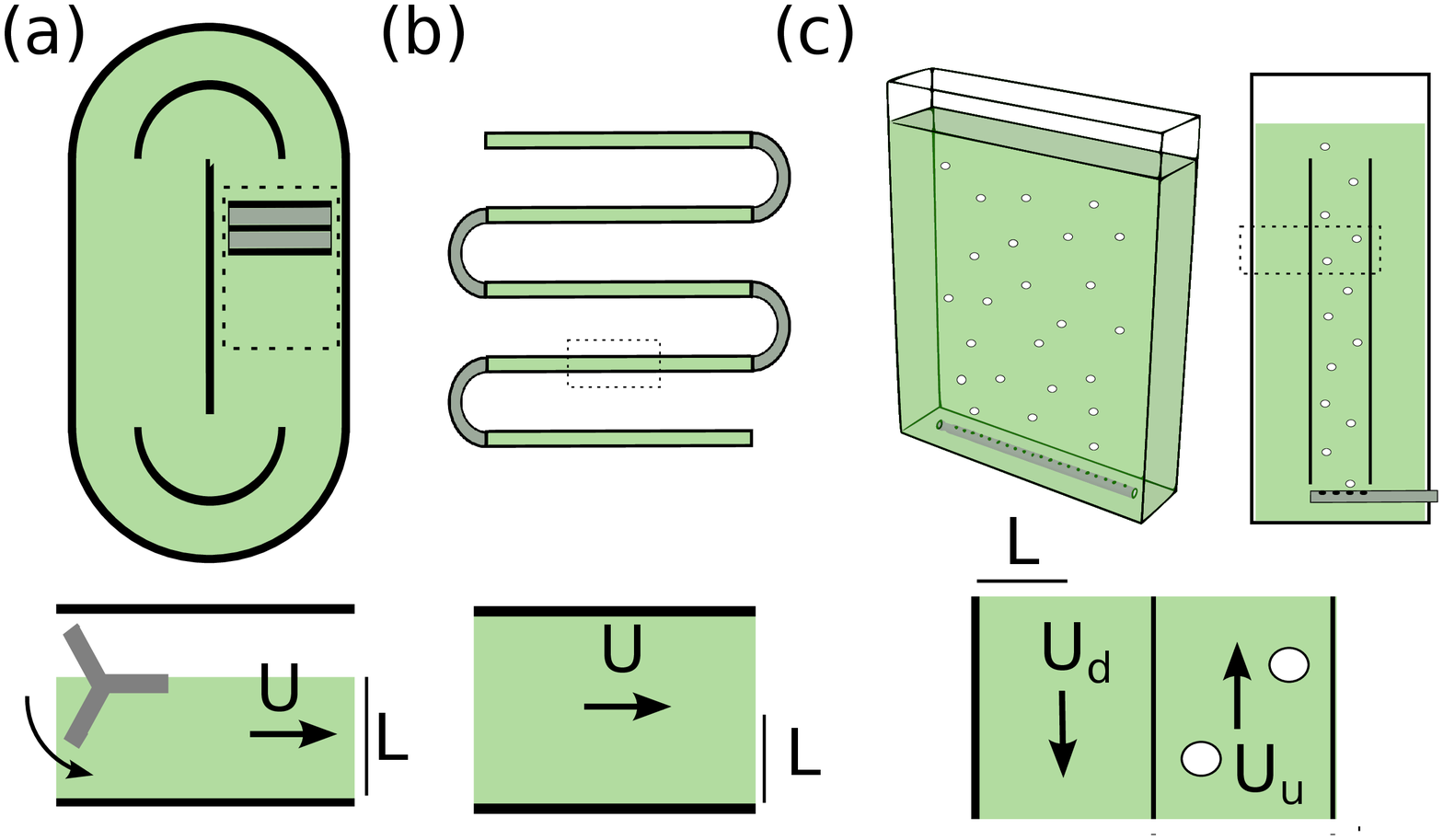}
\caption{Examples of photobioreactors: (a) raceway pond (top view), showing the paddlewheel and guiding baffles; (b) tubular array, where suspensions are typically driven by a pump; (c) air-lift flat panel (3D view) and draft-tube  (cross-sectional side view), where rising bubbles generate flow as well as providing mixing and aeration. The regions highlighted with broken lines in the top diagrams are reproduced below to indicate the length scale $L$ and magnitude $U$ of typical flows. For the airlift in (c), both upwelling $U_u$ and downwelling  $U_d$ flows are shown.}\label{bioreactors}
\end{figure}

Microalgae currently play a prominent role in biotechnology. For example, they are cultured commercially as nutritious fish and crustacean feed in aquaculture, and for high-value products that they can synthesize, such as $\beta-$carotene \citep{Borowitzka99}. Microalgae also hold significant potential for future exploitation; they can provide a sustainable biofuel feedstock that does not compete for arable land with food crops. Despite significant efforts, however, current production systems remain too inefficient for algal biofuels to be commercially viable \citep{Stephensonetal10}. Algal culture commonly occurs in open or closed bioreactors, with production systems employing three stages: cell culture, harvest and downstream processing.  Open bioreactors consist of one or more artificial ponds, quiescent or stirred, where stirring allows to obtain greater concentrations per reactor area \citep{Borowitzka99}. A common design is the raceway pond, in which rotating paddles generate flow (see figure \ref{bioreactors}a). Typically, raceway ponds are shallow (depths $L=10-30$ cm), which aids cell exposure to light. Typically, the flow is turbulent (speeds of $30$ cm/s), facilitating light exposure and gas exchange. Pond bioreactors are relatively cheap to operate, but are limited by their low areal productivity and susceptibility to contamination. This restricts the range of viable species in open pond culture to robust, high-value extremophiles, which grow in prohibitive conditions such as high salinity or low pH. A greater variety of fast-growing algae can be cultured in closed bioreactors, where single species are propagated within sealed transparent containers. Common geometries are flat, cylindrical (columnar or tubular), and annular, see figure \ref{bioreactors}b-c. As for raceways, the constraint of light penetration limits the smallest dimension to $10$ cm. In tubular bioreactors the flow-speed typically is $50$ cm/s, whilst flat panel and columnar/annular set-ups operate at $1$-$10$ cm/s \citep{Morweiseretal09}. Although closed bioreactors allow for rapid growth conditions with small footprints, current designs are relatively expensive to manufacture, operate and maintain, and the algae are susceptible to invasion by other less-useful species.

%

Existing bioreactors operate under both laminar and turbulent flow conditions. The transition between the two flow regimes depends on the ratio of inertial and viscous forces in a fluid. This is expressed by the dimensionless Reynolds number $\rm{Re}=U L/\nu$, where $U$ and $L$ are the characteristic speed and length scale of the flow, respectively, and $\nu$ is the suspension kinematic viscosity \citep{Pope00}. In general, flows with $\rm{Re}>2000$ will be turbulent, though the transition to turbulence depends on geometry and particular flow conditions. Approximating the kinematic viscosity by that of water and using the flow speed and length scales above, we see that flows in air-lift reactors can be both laminar or turbulent (Re$\gtrsim100$), while raceway ponds and tubular reactors are always turbulent (Re$>30000$). The mixing properties of turbulent flows are thought beneficial for cell growth, although clear evidence for this is lacking. With air-lifts the mixing is caused by rising bubbles, which also provide aeration, and the flow does not need to be turbulent. Both cells and nutrients will disperse in the above flows, with cell growth and productivity depending critically on such dispersion.

%
%
%
%
%
%
%
%
%
%
%
%
%
%
%
%
%
%


Current photobioreactor designs assume that cells disperse like tracers, whether they swim or not \citep{CamachoRubioetal04}. However, as detailed above, recent studies show that the distribution and consequent dispersion of swimming cells in flows, including biotechnological interesting species like {\it Dunaliella}, should be very different from that of passive tracers \citep{BeesCroze10, BearonBeesCroze12}. For example, if cells disperse differently to nutrients in a reactor they will separate, which could have catastrophic consequences for growth. Inspired by the possibility of taking advantage of the peculiarities of swimmer response to flow to improve bioreactor operation, we study here the dispersion of gyrotactic swimming algae in laminar and turbulent channel flows. We carry out time-resolved direct numerical simulations, comparing results to predictions from the Bees \& Croze dispersion theory applied to channel geometry. The paper is structured as follows. In Section 2, we present the simulation methods and outline a derivation of the theory for the new geometry. In Section 3, we present simulation results for the dispersion of biased swimming algae in laminar and turbulent flows, comparing theory and simulation results for laminar flows. In Section 4, we interpret the results, and discuss their implications for dispersion in photobioreactors and the environment.

\section{Methods \label{Methods}}

\subsection{Direct numerical simulations (DNS): governing equations}

We simulate the dynamics of a population of biased swimming micro-organisms, typically $N=2 \times 10^5$ and $10^6$ individuals per simulation (see table \ref{table:cases}), placed in laminar and turbulent flows. Each microswimmer is modelled as a spheroidal particle whose position $\mathbf{x}_i$, $i=1..N$, evolves according to
\begin{equation}
\frac {d \mathbf{x}_i}{d t}= \mathbf{u} +  V_s \mathbf{p} 
\end{equation}
where $V_s$ is the mean cell swimming speed, $ \mathbf{u}$ is the local fluid velocity and $\mathbf{p}$ is the local particle orientation.

The orientation $\mathbf{p}$ of each swimmer evolves in response to the biasing torques acting upon it. Making the realistic simplifying assumption that swimming cells have a spheroidal geometry, the reorientation rate of the organisms is defined by the inertia-free balance of gravitational and viscous torques \citep{Jeffery22, PedleyKessler92}
\begin{equation}\label{pp_eq}
\frac {d \mathbf{p}}{d t} = \frac{1}{2 B} \left[  \mathbf{k} -  (\mathbf{k}\cdot \p) \p \right]
+ \frac{1}{2} \xo \times \p + \alpha_0 (\mathbf{I}-\p\p) \cdot \mathbf{E} \cdot \p +\mathbf{\Gamma}_r,
\end{equation}
where $\mathbf{k}$ is a unit vector in the vertical direction; $B={\mu\alpha_{\perp}}/(2h\rho g)$ is the gyrotactic reorientation time ($h$ denotes the centre-of-mass offset relative to the centre-of-buoyancy, $\alpha_{\perp}$ is the dimensionless resistance coefficient for rotation about an axis perpendicular to $\mathbf{p}$, $\rho$ and $\mu$ are the fluid density and viscosity, respectively); $\alpha_0=(a^2-b^2)/(a^2+b^2)$ is the eccentricity of the spheroids with major axis $a$ and minor axis $b$; and, finally, $\mathbf{E}$ is the rate of strain tensor and $\xo$ the vorticity. The noise $\mathbf{\Gamma}_r$ is added to simulate the stochastic rotational diffusivity of a swimming cell. In the simulations it is implemented with random angular steps of magnitude $\sqrt{2d_r}$, so that cells diffuse with rotational diffusivity $d_r$ \citep{LocseiPedley09, ThornBearon10}.

The flow field $\mathbf{u}$ is obtained by solving the Navier--Stokes equations for an incompressible viscous fluid, such that
\begin{equation} \label{NS}
\frac{\rm{D}\mathbf {u} }{\rm{D} t}  =  - \frac{1}{\rho}\nabla p +  \nu \nabla^2 \mathbf {u},\mbox{~~~~~~~~~}
\nabla \cdot  \mathbf {u} = 0,
\end{equation}
where $\frac{\rm{D}}{\rm{D} t}\equiv\frac{\partial}{\partial t} + (\mathbf {u} \cdot \nabla)$ denotes the material derivative. Here, for simplicity, we shall assume cells to be neutrally buoyant, neglecting feedback from the particles to the flow (these effects are explored elsewhere \citep{BeesCroze10}).

\subsection{DNS: geometry, scaling and statistical measures of dispersion \label{DNSstatmeas}}

We consider a standard channel geometry: two flat plates parallel to each other, infinite in extent, and separated by a gap of size $2H$. When focusing on swimming cells it would seem natural to rescale length by the plate half-width $H$ and time by $H^2 d_r/V_s^2$, where $V_s$ is the mean swimming speed and $d_r$ is the cell rotational diffusivity, the characteristic time a cell swimming with $V_s$ takes to diffuse across $H$ with diffusivity $V_s^2 d_r^{-1}$. However, the parameter values obtained from this `cell-based' rescaling are too large and thus numerically inconvenient for simulations. Thus we adopt a `flow-based' rescaling in terms of the characteristic length $H$ and the flow-based timescale $H/U_c$, the time taken for a flow with centerline speed $U_c$ to advect a cell by $H$. In terms of this rescaling, the dimensionless equations of motion for a biased swimming cell are
\begin{eqnarray} \label{NS_dimless1}
\frac {d \mathbf{x}_i^*}{d t^*}&=& \mathbf{u^*} +  v_s \mathbf{p},\\
\frac {d \mathbf{p}}{d t^*}  &=& \eta^{-1} \left[  \mathbf{k} -  (\mathbf{k}\cdot \p) \p \right]
+ \frac{1}{2} \om^* \times \p + \alpha_0 (\mathbf{I}-\p\p) \cdot \mathbf{E^*} \cdot \p +\mathbf{\Gamma}_r^*, \label{NS_dimless2}\\
\frac{\rm{D}\mathbf {u^*} }{\rm{D} t^*}& = & - \nabla^* p + Re^{-1} \nabla^{*2} \mathbf {u}^*, \mbox{~~~~} \nabla \cdot  \mathbf {u}^* = 0,\label{NS_dimless3}
\end{eqnarray}
where starred quantities denote dimensionless variables. In particular, we define the dimensionless swimming speed $v_s= V_s/U_c$, the gyrotactic parameter $\eta= 2 B U_c/H$ and the Reynolds number
\begin{equation} \label{Re}
\rm{Re}=\frac{U_c H}{\nu},
\end{equation}
where $U_c$ is the centerline speed, $H$ is the channel half-width and $\nu$ the fluid's kinematic viscosity. Non-dimensionalising the noise $\mathbf{\Gamma}_r$ also defines the dimensionless rotational diffusivity $d_r^*=d_r H/U_c$. 

We adopt a Cartesian coordinate system where the mean flow in the $x$ (streamwise) direction, varies with the wall-normal coordinate $y$, and is independent of $z$ (spanwise direction). We integrate equations (\ref{NS_dimless1}), (\ref{NS_dimless2}) and (\ref{NS_dimless3}) numerically (see supplementary materials) to find $\mathbf{x}_i^*(t^*)=[x_i^*(t^*),y_i^*(t^*),z_i^*(t^*)]$. Enumerating the number of cells $N(\mathbf{x})$ at a given position $\mathbf{x}$ in a bin of fixed volume $\Delta V$, we obtain the cell concentration $c(\mathbf{x})=N(\mathbf{x})/\Delta V$. 
From the cell coordinates can we further define statistical measures of the cell dispersion in a flow: the adimensional drift $\Lambda_0^*$ with respect to the mean flow $U=(2/3)U_c$; the effective streamwise diffusivity $D^*_e$ and the skewness of the cell distribution, $\gamma$. These measures of dispersion are given by
\begin{equation} \label{statmeas}
\Lambda_0^*( t^*)\equiv\frac {d m_1^*}{d t^*}-\frac{2}{3},\mbox{~~} D^*_e( t^*)\equiv\frac{1}{2}\frac {d}{d t^*}\Var(x^*),\mbox{~~}
\gamma(t^*)\equiv \frac{m_3^* -3m_1^* m_2^* +2 m_1^{*3}}{\Var(x^*)^{3/2}},
\end{equation}
where $m_p^{*}=\frac{1}{N} \sum_i (x_i^*)^p$ ($p=1,2,3$) are the distribution moments and $\Var(x^*)=m_2^{*}-m_1^{*2}$ is the variance of the cell distribution. The statistical measures can be transformed from the flow-based scaling with characteristic time-scale $H/U_c$ to the cell-based scaling with time-scale $H^2 d_r/V_s^2$  by the transformations
\be
t^*\to t \equiv t^*/\rm{Pe},\mbox{~~}  \Lambda_0^*\to \Lambda_0 \equiv \Lambda_0^*\,\rm{Pe}, \mbox{~~}  D_e^*\to D_e \equiv D_e^*\,\rm{Pe},
 \ee
where $\rm{Pe}$ is the cell P\'{e}clet number defined with respect to the center-line speed (see equation (\ref{eq:Pe})). Note that the skewness $\gamma$ does not depend on the scaling used. We shall present results in terms of the cell-based scaling and compare their limit at long times with analytical predictions.


\subsection{Analytical theory: dispersion in laminar flows at long times \label{antheory}}

Here, we shall obtain analytical predictions for the dispersion of algae swimming in laminar channel flow that will be compared with the results from the simulations. The general Bees \& Croze \citep{BeesCroze10} continuum dispersion theory for biased swimmers will be applied to the new channel geometry; it is valid in the long-time limit ($t\gg H^2 d_r/V_s^2$). The derivation for channel flow is similar to the pipe flow example in Bees \& Croze, so we provide an outline here. 
Those readers that are less mathematically inclined may skip the derivation of the results in this section.

We shall begin with the continuity equation for the cell number density $n$:
\be
\frac {\partial n}{\partial t} =  - \nabla \cdot \left[ n \left ( {\bf u } + {\bf V}_c
\right) - {\bf  D} \cdot \nabla  n \right]. \label{eq1:m3}
\ee
The flow velocity $\mathbf {u}$ is the solution of the Navier-Stokes equation (\ref{NS}). We adopt the same coordinate system described above for the DNS. For laminar flows such that the flow downstream is only a function of the wall-normal coordinate $y$, the flow velocity can be expressed as
\be
\xu(y) = u(y) \ex = U[1+\chi(y)] \ex, \label{eq:mf}
\ee
where $U$ is the mean flow speed and $\chi(y)$ describes the variation in the streamwise direction about this mean. It is convenient to translate to a reference frame travelling with the mean flow, and non-dimensionalise lengthscales by $H$ and timescales by the time to diffuse across the channel $H^2 d_r/V_s^2$. Thus, $\hx = (x-Ut)/H$, $\hat{y} = y/H$, $\hat{z} = z/H$ and $\htt= V_s^2 t /(H^2 d_r)$, where hats denote dimensionless variables.

We assume unidirectional coupling of the cell dynamics to the flow (for bidirectional coupling due to non-neutrally buoyant cells see \citep{BeesCroze10}): cells are biased by shear in the flow, so that the cell swimming velocity ${\bf V}_c= V_s \xq$, where $\xq$ is the mean swimming direction, and diffusivity tensor ${\bf D}$ are functions of local flow gradients.  Here, we consider analytical predictions for spherical cells ($\alpha_0=0)$, so cell orientation is only a function of vorticity, $\mathbf{\omega}=\nabla\times \mathbf{u}=-\chi^\prime {\bf e}_z$.

Consider planar Poiseuille flow, $\chi=(1-3y^2)/2$ and $U=(2/3)U_c$ \citep{Wooding60}, where $U_c$ is the centerline (maximum) flow speed.
Equation (\ref{eq1:m3}) becomes
\be
\frac {\partial n}{\partial t}  = \nabla \cdot \left( \xD \cdot \nabla n \right) - \Pem \chi n_{x}
- \beta \nabla \cdot \left( n \xq \right), \label{CC1}
\ee
with
\be
\Pe = \frac{U_c H d_r}{V_s^2}, \mbox{~~~~and~~~~}
\beta = \frac{H d_r}{V_s},
\label{eq:Pe}
\ee
where hats have been dropped for clarity. Pe is a P\'{e}clet number, the dimensionless ratio of the rates of transport by the flow and swimming diffusion, and $\beta$ is the ratio of channel half-width to the length a cell swims before reorienting significantly. Alternatively, we can re-write $\beta=H V_s^2 d_r/V_s^2$ and interpret it as a `swimming' P\'{e}clet number, the ratio of the rates of transport by swimming and diffusion. No-flow and no-flux boundary conditions will be applied to (\ref{CC1}), such that
\be
\xu = {\bf 0} \mbox{~~~~and~~~~}
\xn \cdot \left( \xD \cdot \nabla n - \beta \xq n \right) = 0, \mbox{~~~~on~~~~} \Sigma ,
\label{nfnf}
\ee
where $\xn$ is the unit vector normal to the channel boundary $\Sigma$.

As the flow is translationally invariant along $x$, the mean swimming direction and diffusivity tensor are independent of $x$: $\xD=\xD(y)$, $\xq=\xq(y)$. This permits a treatment of dispersion using moments in a similar vein to that of Aris \citep{Aris56}. The $p$th moment with respect to the axial direction is defined as
\be
c_p(y,t) = \int^{+\infty}_{-\infty} x^p n(x,y,t) dx,
\ee
provided $x^p n(x,y,t)\to 0$ as $x\to\pm\infty$. We denote cross-sectional averages by overbars. The cross-sectionally averaged axial moment is thus
\be
m_p(t) =  \overline{c_p} = \frac12\int_{-1}^1 c_p(y,t) dy.
\ee

 In Cartesian coordinates equation (\ref{CC1}) becomes
 \bea\label{CC2}
 n_{t} &=&(D^{yy}n_y-\beta q^y n +D^{xy} n_x)_y+D^{x y} n_{x y}  \\ \nonumber
 & & -  [\Pem \chi + \beta q^{x}] n_{x}+D^{x x} n_{x x}.
 \eea
 Multiplying by $x^p$ and integrating over the length of an infinite channel we obtain the moment evolution equation
 \bea
 c_{p,t} &=& (D^{yy}c_{p,y}-\beta q^y c_p -p D^{xy} c_{p-1})_y-p D^{x y} c_{p-1,y}  \\ \nonumber
 & &+ p [\Pem \chi + \beta q^{x}] c_{p-1}+p(p-1) D^{x x} c_{p-2}.
 \eea
 Averaging over the cross-section and applying the no-flux boundary conditions
(\ref{nfnf}), yields
 \be
 m_{p,t} = p(p-1)\overline{D^{x x}c_{p-2}} -  p \overline{D^{x y} c_{p-1,y}}
 + p \overline{\left[ \Pem \chi + \beta q^{x} \right] c_{p-1} }\label{B}
 \ee
from which we calculate measures of cell dispersion. The drift above the mean flow,
$\Lambda_0\equiv\lim_{t\to\infty}\frac{d}{d t}m_1(t)$, is given by
\be\label{drift}
\Lambda_0 =  -\overline{D^{xy}Y_{0}^{0\prime}}
+ \overline{\left[ \Pem \chi + \beta q^{x} \right] Y_{0}^0},
\ee
where
\be\label{conceq}
Y_0^0(y) = \exp{\left( \beta \int^y_0 \frac{q^y(s)}{D^{yy}(s)}ds \right)}\left\{\overline{ \exp{\left( \beta \int^y_0 \frac{q^y(s)}{D^{yy}(s)}ds \right)}}\right\}^{-1}
\ee
is the zeroth axial moment (normalised concentration profile). Similarly the effective diffusivity, $D_e\equiv\lim_{t\to\infty}\frac{1}{2}\frac{d}{d t}\left[m_2(t)-m_1^2(t)\right]$, is given by:
\be\label{diff}
D_{e}= - \overline{D^{xy} g^{\prime}}
+ \overline{ \left[ \Pem \chi +\beta q^{x} -\Lambda_0\right] g}+ \overline{D^{xx} Y_0^0},
\ee
where $g(y)=Y_0^0 \int^y_0  ( \frac{D^{xy}(s)}{D^{yy}(s)}-  \frac{\tilde{\Lambda}_0(s) - \Lambda_0 \tilde{m}_0(s)}{D^{yy}(s)Y^0_0(s)}) ds$, with $\tilde{\Lambda}_0(y) = \int^y_0  [-D^{xy} {Y_0^0}^{\prime} + \left[ \Pem \chi + \beta q^x \right] Y^0_0]ds$ and $\tilde{m_0}(y) = \int^y_0 Y^0_0 ds$. The weighting function $Y_0^0(y)$ controls the drift and $g(y)$ (related to the first axial moment) controls the value of the diffusivity.

To make predictions from (\ref{drift}) and (\ref{diff}) we require expressions for $\xD$ and $\xq$ from microscopic models of the statistical response of cells to flow. Two main models have been proposed for biased swimming algae: the Fokker-Planck (FP) model \citep{PedleyKessler90, Beesetal98} and generalized Taylor dispersion (GTD) \citep{HillBees02, ManelaFrankel03, BearonBeesCroze12}. For spherical cells, each model predicts how nondimensional swimming direction ${\bf q}$ and diffusivity ${\bf D}$ depend on two nondimensional quantities: $\sigma(y)=-\chi'(y)U/(2 d_r H)=-\chi'(y)(1/3)(\rm{Pe}/\beta^2)$, the ratio of reorientation time by rotational diffusion to that by shear (vorticity), and $\lambda = 1/(2 d_r B)$, the ratio between the time of reorientation by rotational diffusion $(1/2)d_r^{-1}$ and the gyrotactic reorientation time $B$. The functional forms for the transport parameters ${\bf q}(\sigma)$ and ${\bf D}_m(\sigma)$, where subscripts $m=$F and G denote the FP and GTD models, respectively, are given in appendix A. In particular, the two models differ qualitatively in their predictions for the diffusivity as a function of the shear rate and therefore they provide different predictions for cell distribution and dispersion. The dispersion predictions (\ref{drift}) and (\ref{diff}) were evaluated numerically with Matlab (Mathworks, Natick, MA) using ${\bf q}(\sigma)$ and ${\bf D}_m(\sigma)$ from the two models, see supplementary materials.


\subsection{Parameters, simulation time and flow profiles}

In simulations and theoretical evaluations we used the following cell parameters based on {\it C. augustae}: $V_s=0.01$ cm s$^{-1}$ (swimming speed), $d_r=0.067$ s$^{-1}$ (cell rotational diffusivity), and $B=3.4$ s (gyrotactic reorientation time) \citep{BeesCroze10}. Further, we consider the flow of suspensions taken to have the same viscosity as water, $\nu=0.01$ cm$^2$s$^{-1}$. The cell eccentricity $\alpha_0$ is also held fixed. The analytical theory presented above assumes that cells are spherical, $\alpha_0=0$, but laminar and turbulent simulations have also been performed for elongated cells with $\alpha_0=0.8$. With these parameters fixed, choosing the centreline speed $U_c$ and channel width $H$ gives the nondimensional flow-based parameters used in the DNS: $v_s=V_s/U_c$ (dimensionless swimming speed), $\eta= 2 B U_c/H$ (gyrotactic parameter), $d_r^*=d_r H/U_c$ (dimensionless rotational diffusivity), and Reynolds number Re$=U_c H/\nu$. For the test run with passive tracers an additional noise term was added to equation \ref{NS_dimless1} to simulate the translational diffusivity $D_t$, nondimensionalised as $D_t^*=D_t/(U_c H)$. These flow-based parameters can be transformed into cell-based nondimensional parameters for comparison with analytical predictions. From the definitions above and equation (\ref{eq:Pe}) it can be shown that Pe$=d_r^*/v_s^2$ (flow P\'{e}clet number), $\beta=d_r^*/v_s$ (swimming P\'{e}clet number) and $\sigma(y)=-(1/3)\chi'(y)/d_r^*$ (local dimensionless shear-rate). Since $\chi'(y)=-3 y$ for plane Poiseuille flow, the maximum dimensionless shear is given by $|\sigma_{max}|=1/d_r^*=\eta\lambda$, where we recall $\lambda=1/(2 B d_r)$, the nondimensional bias parameter. Simulations are more readily interpreted and compared to analytical theory in terms of these parameters (shown in table~\ref{table:cases}).
%
\begin{table}
\centering
{\begin{tabular}{ c| c| c| c c c c c c  }
Re (Lam/Turb)  & Pe &$\beta$ & $\alpha_0$ \\
\hline
 4 (L)  & 27 & 1.34, 6.7, 33.5 & 0 \\
100  (L) & 670  & 1.34, 6.7, 33.5 & 0 (for all $\beta$), 0.8 (for $\beta=6.7$)  \\
250 (L)  & 1675 & 3.35, 33.5 & 0\\
2500 (T) & 16750  & 33.5 & 0\\ 
4200 (T) & 28140   & 55 & 0, 0.8\\ 
10000 (T) & 67000  & 67 & 0 
\end{tabular}}
\caption{Nondimensional parameters for the different cases considered in the DNS and in the analytical model predictions for the laminar case. Re is the Reynolds of the flow, Pe and $\beta$ are flow and swimming P\'{e}clet numbers, respectively, defined by equations (\ref{eq:Pe}), and $\alpha_0$ is the cell eccentricity ($>0$ for elongated cells, zero for spheres).}
\label{table:cases}
\end{table}

\begin{figure}[h!]
\centering
\includegraphics[width=84mm]{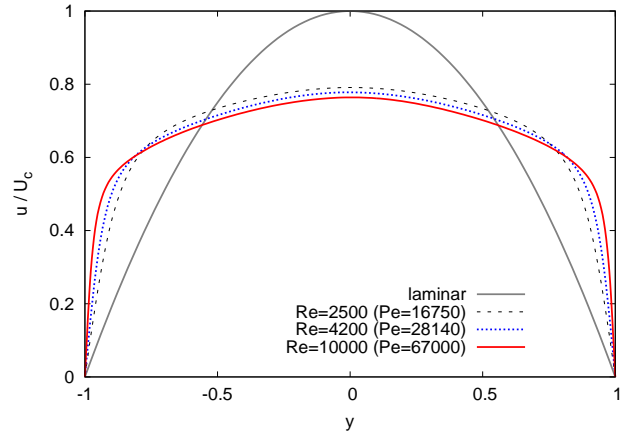}
\caption{Flow profiles for the dimensionless flow speed as a function of $y$ for the laminar cases (self-similar for all Re) and time-averaged flow profiles for the turbulent cases (Re, Pe)$=(2500,16750)$, ($4200,28140$) and ($10000,67000$), as shown.}\label{mean_flow}
\end{figure}

The dimensionless flow profiles corresponding to the Reynolds numbers used in this study are plotted in figure \ref{mean_flow} for the benefit of the reader. Laminar flows are self-similar, so the nondimensional flow has the same parabolic profile for all Re. Time-averged turbulent flows have distinct profiles that depend on the Reynolds number. Note that the number of degrees of freedom in the turbulent flow simulations scales as Re$^9$ \cite{Pope00}. This makes large Re simulations computationally expensive. For this reason we do not investigate dispersion for flows beyond Re$=10000$.

\section{Results \label{Results}}

\subsection{Passive tracers: classical dispersion \label{passive}}

\begin{figure}[tbh!]
\centering
\includegraphics[width=84mm]{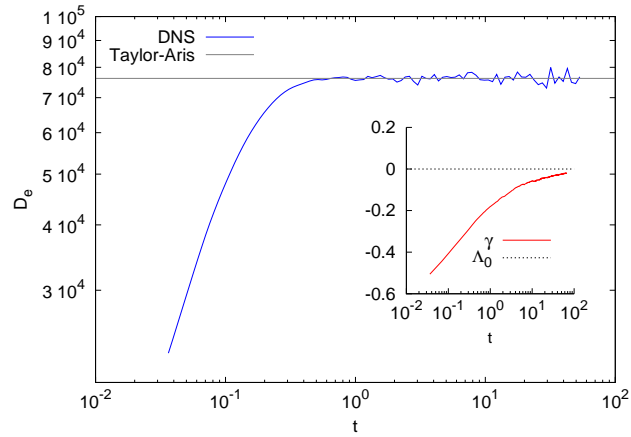}
\caption{Time dependence of the effective diffusivity, $D_e$, for passive tracers in a laminar flow with Pe$=3000$. For long times $D_e\simeq 76191$, the constant value predicted from the classical Taylor-Aris dispersion for passive tracers (see text). The inset shows the expected zero drift, $\Lambda_0$, above the mean flow and the skewness, $\gamma$, tending to zero at long times, suggesting an approach to the expected normally distributed axial concentration profile \citep{Aris56}.}\label{measures_passive}
\end{figure}

The dispersion of passive tracers, such as molecular dyes or nonmotile cells, is generally well understood. In laminar channel flow passive tracers are transported on average at the mean flow speed; there is no drift relative to mean flow: $\Lambda_0=0$. The effective axial diffusivity $D_e$, is given at long times by the Taylor-Aris result $D_e=1+(8/945)$Pe$^2$ \citep{Wooding60, BrennerEdwards93}. As a benchmark test, we carried out direct numerical simulations for passive tracers (solving equation (\ref{NS_dimless1}) with $v_s=0$ and
 translational noise to simulate molecular diffusivity, see methods). A typical result is shown in figure \ref{measures_passive} for Pe$=3000$. As expected at long times $\Lambda_0\to0$ and the Taylor-Aris diffusivity prediction, $D_e=76191$, compares very well with simulation results. The skewness slowly tends to zero at long times as $t^{-0.5}$ (see supplementary material), suggesting approach to a normal distribution \citep{Aris56}.
The above results for passive tracers will be compared with the dispersion of gyrotactic swimmers. They can also be used to check our analytical theory in the limiting case of unbiased swimmers with no gyrotactic response to flow. In this limit, ${\bf q}={\bf 0}$ and the diffusivity tensor is isotropic ${\bf D}={\bf I}/6$, where ${\bf I}$ is the identity matrix (see also \citep{BearonBeesCroze12}). Thus, in the calculation of equations (\ref{drift}) and (\ref{diff}), $D^{xx}=1/6=D^{yy}$, $q_i=0$ and $D^{xy}=0$ (as $Y_0^0=1$ and $J(y)=0$), so that $\Lambda_0=0$, as expected for passive tracers, and $D_e=1/6+6(8/945)$Pe$^2$.  In the Bees \& Croze theory $D_e$ and Pe are scaled with respect to the diffusivity scale $V_s^2/d_r$. Using the scaling $(1/3)V_s^2/(2 d_r)$ for random diffusers in 3D, we recover $D_e^\prime=1+(8/945)$Pe$^{\prime 2}$, the classical Taylor-Aris result for channels.

The dispersion of passive tracers in turbulent channel flows has also been elucidated \citep{Fisher73}.
Elder derived the (dimensional) effective diffusivity $K$ of passive tracers in turbulent open channel flow as $K=5.9\,u_{\tau} H$, where $u_{\tau}$ is the friction velocity and $H$ is the channel depth \citep{Elder59, Fisher73}. This open channel result applies equally to a closed channel with half-width $H$. Using the approximation $U_c/u_{\tau}\approx 5 \log_{10}$Re \citep{Pope00}, this result can be written as $K\approx2.72\nu$Re$/\ln($Re$)$. To compare results across the laminar-turbulent transition (see later in figure \ref{disp_lamturb}), we use equations (\ref{Re}) and (\ref{eq:Pe}) to obtain a nondimensional turbulent diffusivity $D_e=K/D_0$ as a function of Pe, where $D_0$ is the characteristic diffusivity scale. However, we stress that the diffusivity depends only on Re in the turbulent case: molecular diffusivity is negligible  \citep{Taylor54a}. Notice also from the result of Elder how the effective turbulent diffusivity grows less sharply with Pe than in the laminar regime. Below, we will contrast the classical predictions for dispersion discussed in this section with our new results for biased swimmers.

\subsection{Gyrotactic algae: a new, non-classical `swimming dispersion'}


\subsubsection{Laminar downwelling flows: gyrotactic swimming strongly affects dispersion \label{lamdown}}

\begin{figure}[tbph!]
\centering
\includegraphics[width=84mm]{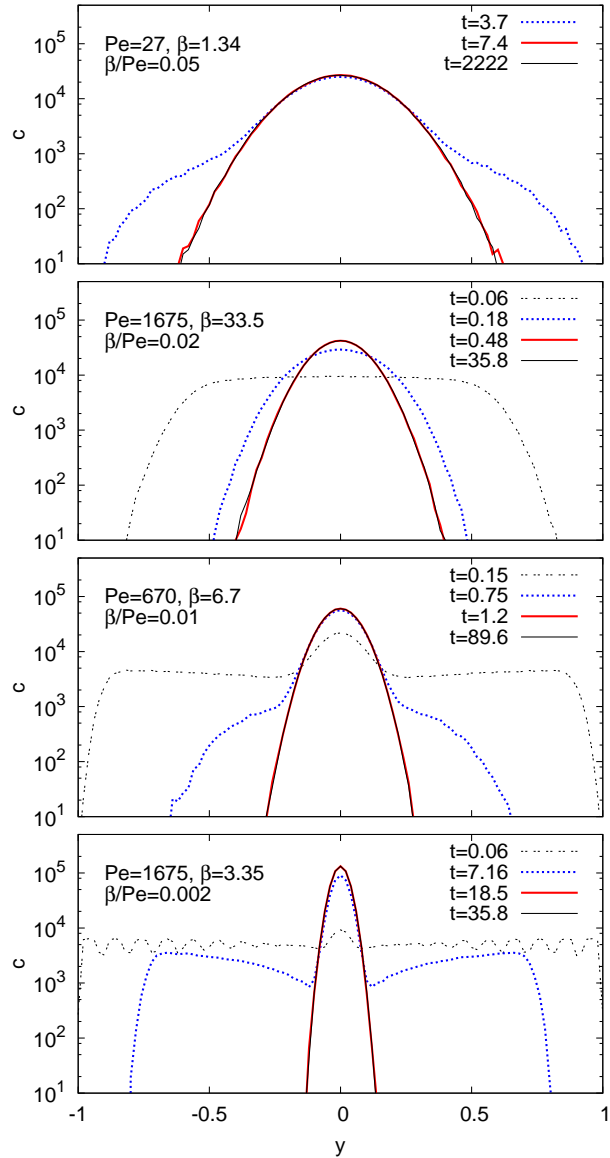}
\caption{Evolution of the DNS concentration profiles (not normalised) in the wall normal direction $y$ for gyrotactic cells in downwelling channel flow with $($Pe, $\beta)$=$(27,\,1.34)$, $(1675,\,33.5)$, $(670,\,6.7)$, $(1675,\,3.35)$, from top to bottom as shown. At large times, the gyrotactic cells are observed to accumulate around the channel center. The profiles become more peaked as the ratio $\beta/$Pe is decreased, see figure \ref{steady_prof}.}\label{conc_prof_lam}
\end{figure}

It is illuminating to consider how gyrotactic cells distribute across a channel in downwelling laminar flows, as this determines their streamwise dispersion. Figure \ref{conc_prof_lam} displays the time-evolution of the cross-sectional cell concentration profiles in the wall-normal direction $y$ for a selection of values of $($Pe, $\beta)$. It is seen that an initially uniform concentration profile evolves to one focused at the channel centre. This is what one would expect for swimming gyrotactic cells with orientations biased by combined gravitational and viscous torques \citep{Kessler85, PedleyKessler92}.
At the population scale, the long-time concentration distribution is a result of the balance between cross-stream cell diffusion and biased swimming. In section (2\ref{antheory}), the cell concentration (normalised by its mean) is given by equation (\ref{conceq})): $Y_0^0=c(y)/\bar{c}=\exp{\left( \beta \int^y_0 (q^y/D_m^{yy})ds \right)}$. (Recall subscripts $m=$G, F denote solutions of the GTD and FP microscopic models respectively.) The two models predict a qualitatively different dependence of concentration distribution on Pe and $\beta$. 
Whilst the FP approach allows for cells to diffuse more and focus less as the cells tumble in flows with large shear rates, the GTD model finds that with increasing local shear rate, $\sigma$, the diffusivity of tumbling cells decreases faster than the decrease in cross-stream cell focusing.
\begin{figure}[tbph!]
\centering
\includegraphics[width=84mm]{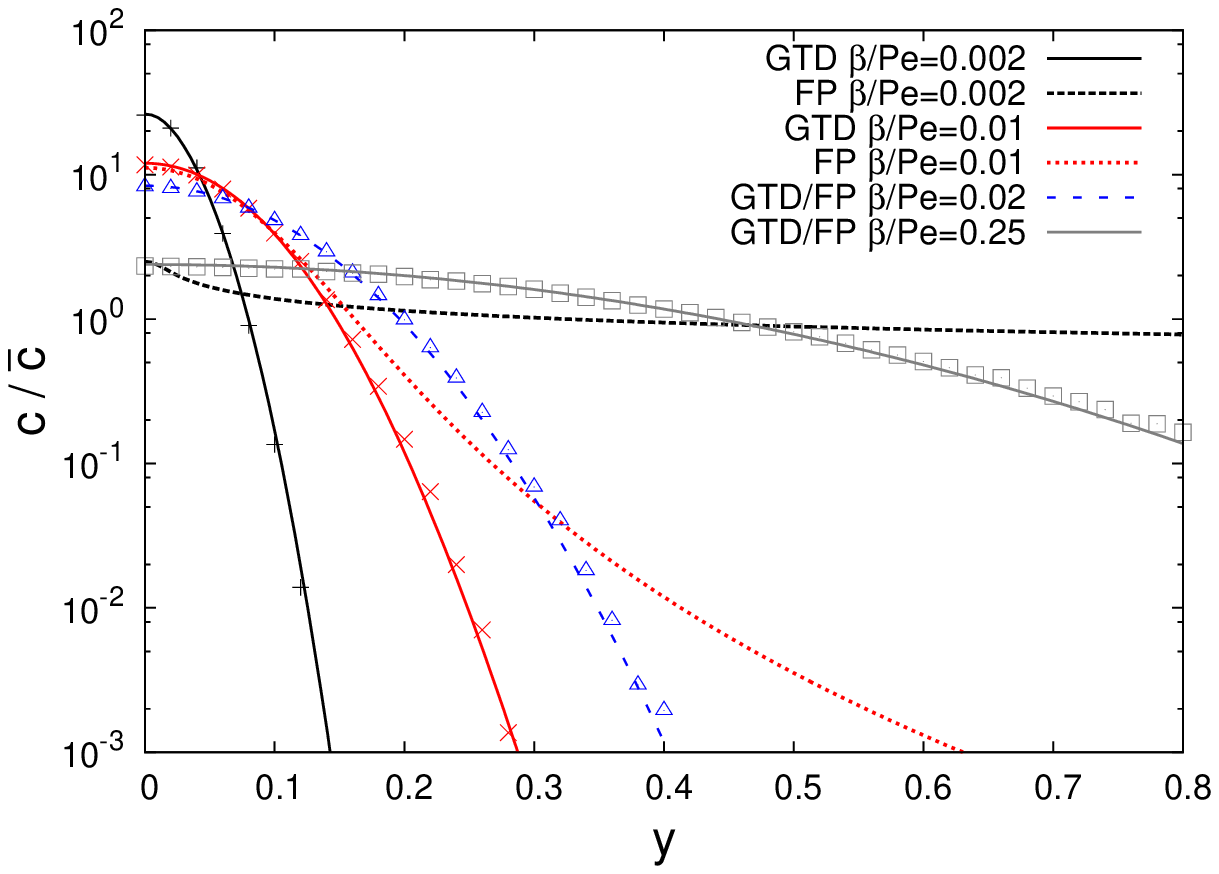}
\caption{Long-time concentrations profiles from DNS, normalised by the mean concentration $\bar{c}$, with $($Pe, $\beta$, $\beta/$Pe$)$: $(1675,\,3.35,\,0.002)$ ($+$), $(670,\,6.7,\,0.01)$ ({\color{red}$\times$}), $(1675,\,33.5,\,0.02)$ ({\color{blue}$\triangle$}), $(27,\,6.7,\,0.25)$ ({\color{grey}$\square$}). Theoretical predictions from the GTD and FP models (lines, see legend), are compared with the simulations. GTD predictions agree very well with DNS profiles for all $\beta/$Pe, but for FP agreement is poor for small values of this ratio. As in figure \ref{conc_prof_lam}, DNS profiles broaden with increasing $\beta/$Pe is increased. This is as predicted by GTD, where profiles can be approximated as Gaussians of width $y_0=[(2/\lambda)(\beta/\rm{Pe})]^{0.5}$, see text and appendix A.}\label{steady_prof}
\end{figure}

It has hitherto been unclear if GTD, mathematically more complicated than the FP approach, provides more accurate predictions. Thus we compare the DNS results and theoretical predictions from the two models. Figure \ref{steady_prof} displays the long-time cell distributions from DNS contrasted with theoretical predictions using GTD and FP. It is clear that DNS profiles get broader with increasing $\beta/$Pe, so profiles are shown for different values of this ratio (individual values of Pe and $\beta$ are shown in the figure caption). GTD and DNS profiles agree very well for all the values of $\beta/$Pe simulated. The FP model only agrees well with the DNS for large values of $\beta/$Pe, where predictions coincide with those of GTD. The agreement is poor at low values (very poor for the sharply focused distribution $\beta/$Pe$=0.002$). It is possible to quantify the scaling of  the breadth of the profiles with the ratio $\beta/$Pe. From asymptotic expressions for the ratio $q^y/D_G^{yy}$ \citep{BearonBeesCroze12}, it follows that GTD profiles can be conveniently approximated by Gaussian distributions of width $y_0=[(2/\lambda)(\beta/\rm{Pe})]^{0.5}$ (see appendix A). Clearly these predictions need also to be tested experimentally (see discussion), but the DNS results indicate that GTD is likely a superior model of gyrotactic response to flow than the FP approach.
\begin{figure}[h!]
\centering
\includegraphics[width=84mm]{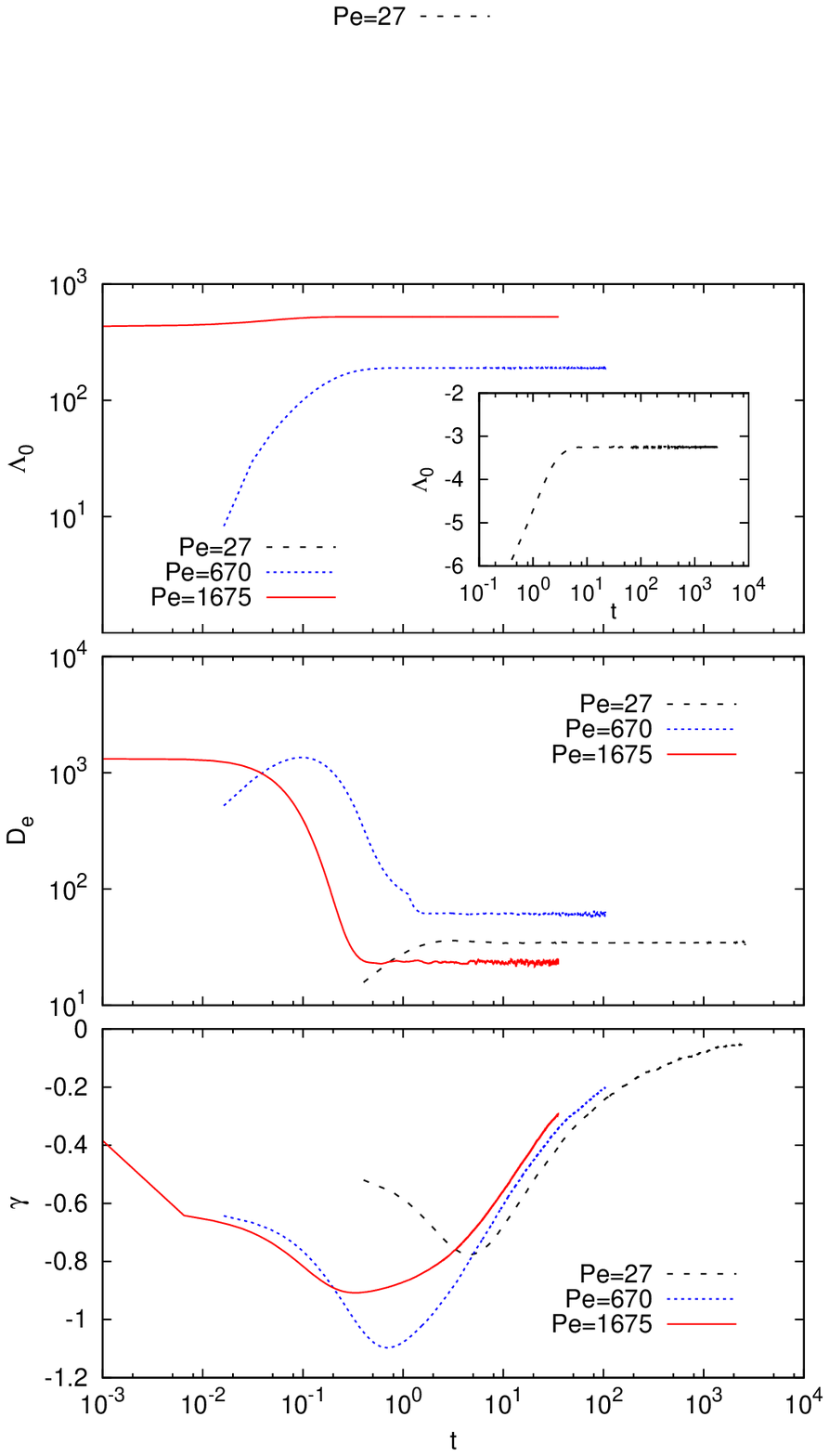}
\caption{Evolution of the dispersion of a population of gyrotactic swimming algae in a vertical down-welling laminar channel flow. The drift above the mean flow, $\Lambda_0(t)$, effective diffusivity, $D_e(t)$, and skewness, $\gamma(t)$ are displayed for Pe$=27, 670$ and  $1675$, top to bottom as shown, for fixed $\beta=33.5$. Uniquely, gyrotactic swimming algae have a non-zero drift above the mean flow, in contrast to the passive tracers in figure \ref{measures_passive}. Even more peculiarly, $\Lambda_0<0$ for Pe$=27$. This is because of cell upswimming at mid-channel, see text.}\label{stat_meas_lam}
\end{figure}

As for the passive case of figure \ref{measures_passive}, we have quantified dispersion using the statistical measures: $\Lambda_0(t)$, the drift above the mean flow;  $D_e(t)$, the effective diffusivity;  and $\gamma(t)$, the skewness. These are plotted in figure \ref{stat_meas_lam} for Pe$=27$, $670$ and $1675$ at the fixed value of $\beta=33.5$. In these simulations, statistically stationary values for $\Lambda_0(t)$ and $D_e(t)$ are achieved for dimensionless times $t\sim1$; steady values are reached earlier for larger Pe. In terms of the Bees \& Croze dispersion theory, steady dispersion is achieved in the long-time limit when transient solutions to the moment equations have died down: $t\gg\tau_1$. The analysis of transient solutions with DNS will be carried out in a future study, but it is reasonable that gyrotaxis makes the approach to steady dispersion faster than for passive tracers. The skewness is negative and approaches zero (with $\gamma\sim t^{-0.49}$ for Pe$=27$; see supplementary materials), suggesting a distribution tending to normality. Bees \& Croze predicted the power law decay $\gamma=\gamma_0 t^{-1/2}$ (the pre-factor $\gamma_0(\beta,$Pe$)$ depends on gyrotactic swimming) with the same exponent as the passive case \citep{Aris56}.

The steady gyrotactic swimmer dispersion displays some very surprising features, evident in the data presented in figure \ref{stat_meas_lam}. For example, $\Lambda_0$, zero in the passive case, grows from a negative value to large positive values as Pe is increased. The effective diffusivity $D_e$, on the other hand, shows a non-monotonic behaviour for increasing Pe (recall $D_e\sim $Pe$^2$ for passive tracers). We can qualitatively account for this behaviour considering the concentration distributions analysed above. Cells are biased to swim to the centre of the channel. Here only the torque due to gravity acts on cells, so cells swim upward at their mean swimming speed, which may be comparable with the mean flow speed for small Pe leading to $\Lambda_0<0$. For large Pe, the upwards swimming speed is negligible.  As cell accumulation at the centre of the channel increases with Pe due to gyrotaxis they drift more relative to the mean flow, and so $\Lambda_0$ is an increasing function of Pe. Cell accumulation at the centre of the channel removes them from regions of high shear rate, eventually leading to a decrease in their effective axial diffusivity with increasing Pe.

The Bees \& Croze dispersion theory allows for the quantification of this intriguing dispersion behaviour; the results are summarized in section (3\ref{longdisp}) below. Prior to this, we shall consider time-dependent dispersion in turbulent flows.

\subsubsection{Turbulent downwelling flows: persistent but weaker swimming dispersion \label{turb_down}}

Here we describe the first DNS study of the dispersion of gyrotactic cells in turbulent channel flows. To compare turbulent and laminar results we first focus on downwelling flows. Simulations were performed for (Pe, Re)$=(16570, 2500$), ($28140, 4200$) and ($67000, 10000$), with corresponding values of $\beta=33.5$, $55$ and $67$.
\begin{figure}[h!]
\centering
\includegraphics[width=84mm]{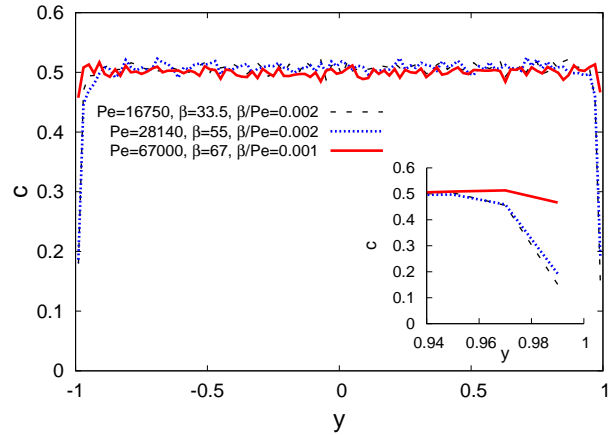}
\caption{Single realizations of the long-time concentration profiles (not normalised) in direction $y$ for gyrotactic cells in downwelling turbulent flows for (Pe, Re)$=(16570, 2500$), ($28140, 4200$) and ($67000, 10000$), as shown. Gyrotaxis causes depletion of cells from regions close to the walls where shear is large, but the mean profile departs little from uniform concentration (expected of passive tracers).}\label{turb_conc}
\end{figure}
The long-time wall-normal cell concentration profiles for these turbulent flows are shown in figure \ref{turb_conc}. It is clear that the mean cell concentration is uniform (barring small fluctuations about the mean) for almost the entire channel width, except for regions close to the wall that are gyrotactically depleted of cells (gyrotactically depleted regions occupy only a small fraction ($< 4\%$) of the channel width). Shear and advection experienced by a cell in turbulent flows are very different from the laminar case. The turbulent flow can be thought of as a time-averaged base profile on which are superposed turbulent fluctuations (the well-known Reynolds decomposition \citep{Pope00}). The shear rate of the base profile is close to zero in the middle of the channel, and large at the walls, see figure \ref{mean_flow}, and deviates from the laminar case; this alone will lead to broader concentration profiles. On top of this, turbulent fluctuations perturb the flow, causing cell reorientation and advection. We can think of turbulence as endowing cells with an increased diffusivity \citep{Lewis03} acting to make gyrotactic accumulations in this downwelling case less pronounced.  Only close to the walls is the impact of mean shear sufficient to cause significant cell depletion; an effect that increases with $\beta/$Pe in a similar fashion to the laminar case.
%
\begin{figure}[h!]
\centering
\includegraphics[width=84mm]{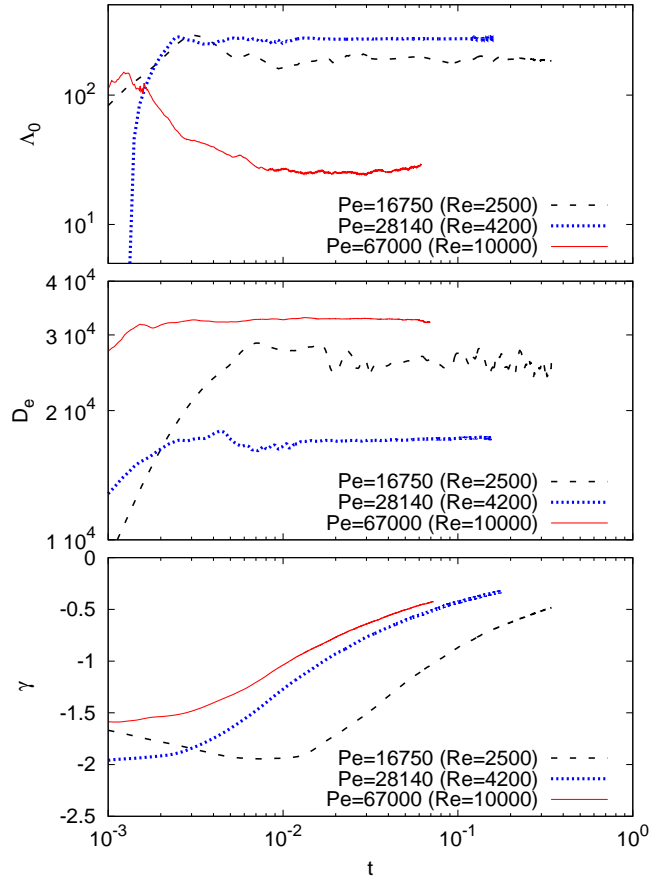}
\caption{Evolution of statistical measures (top to bottom, as for the laminar case) for gyrotactic cells in a vertical downwelling turbulent channel flow for (Pe, Re)$=(16570,2500)$, ($28140$, $4200$) and ($67000$, $10000$), as shown. The long-time values of the drift above the mean flow, $\Lambda_0$ and effective diffusivity, $D_e$ are plotted in figure \ref{disp_lamturb}. Values of $\beta$ are not shown for clarity, see table \ref{table:cases}.}\label{turb_gyro}
\end{figure}
The small but measurable effects of gyrotaxis on the concentration profiles are reflected in the time-dependent dispersion measures for the turbulent case, plotted in figure \ref{turb_gyro} for the same values of $($Pe, $\beta)$ considered  above. We leave a detailed analysis of transients to a future study but note that, due to the increased
diffusivity by turbulence, the approach to the limiting behaviour is faster than in the laminar case. 
Less obviously, the long-time dispersion retains a rich behaviour (observe the order of the curves in figure \ref{turb_conc}). In particular, notice that cells have a nonzero drift $\Lambda_0$; this is due to local focusing of cells in downwelling regions of the fluctuating flow. The dispersion of gyrotactic swimmers is thus qualitatively distinct from that of passive tracers even in a turbulent flow. As Pe is increased to the maximum value simulated, $\Lambda_0$ decreases whilst $D_e$ increases, indicative of increased mixing by turbulence.

\subsubsection{Dispersion in upwelling flows}

\begin{figure}[h!]
\centering
\includegraphics[width=84mm]{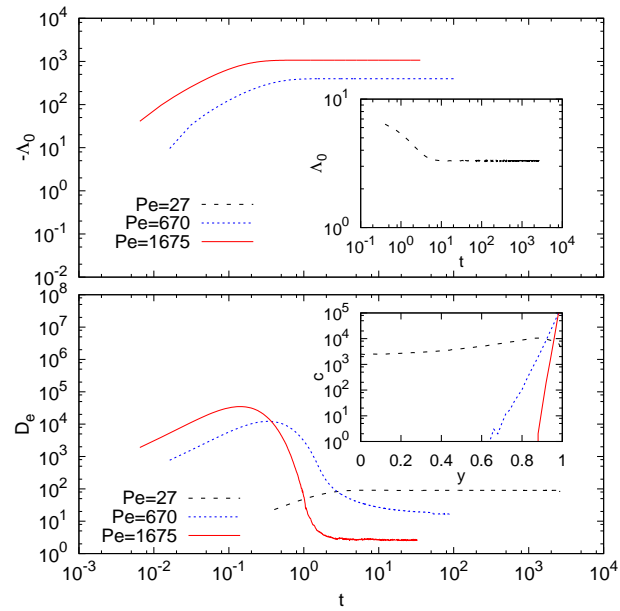}
\caption{Dispersion of gyrotactic algae in vertical upwelling laminar flows. The parameters are as for the downwelling laminar case ($\beta=33.5$ and $Pe=27, 670, 1675$). Top: the drift $-\Lambda_0$ (inset: the positive drift for $Pe=27$). Bottom: the effective diffusivity $D_e$ (inset: the normalised concentration profiles).}\label{upwelling}
\end{figure}

The case of dispersion in flows directed vertically upwards (against the direction of gravity) is considered here briefly. The DNS results for the same values of Pe and $\beta$ as the laminar downwelling case of section (3\ref{lamdown}) are shown in Figure \ref{upwelling}. The drift above the mean flow $\Lambda_0$ is positive for small Pe, and grows more negative with increasing Pe.
%
This behaviour is the result of the peculiar distribution of gyrotactic cells in the flow:
cells in upwelling flow are biased to swim not to the channel centre, but to the walls \citep{Kessler85}. Interestingly, accumulation and dispersion depend critically on the flow direction! The inset of figure \ref{upwelling}, bottom, shows the normalised cell concentration $c/\bar{c}$ in the wall normal direction $y$, demonstrating the wall accumulation, which becomes more peaked with decreasing $\beta/$Pe (for more details, see supplementary materials). Thus, for fixed $\beta$ and increasing Pe, cells  focus at the walls and experience slower flow and less of the shear profile, leading to a decrease in both the drift, $\Lambda_0$, and the diffusivity, $D_e$. The change of sign in $\Lambda_0$ has a similar explanation as for the downwelling case.
%
\begin{figure}[h!]
\centering
\includegraphics[width=84mm]{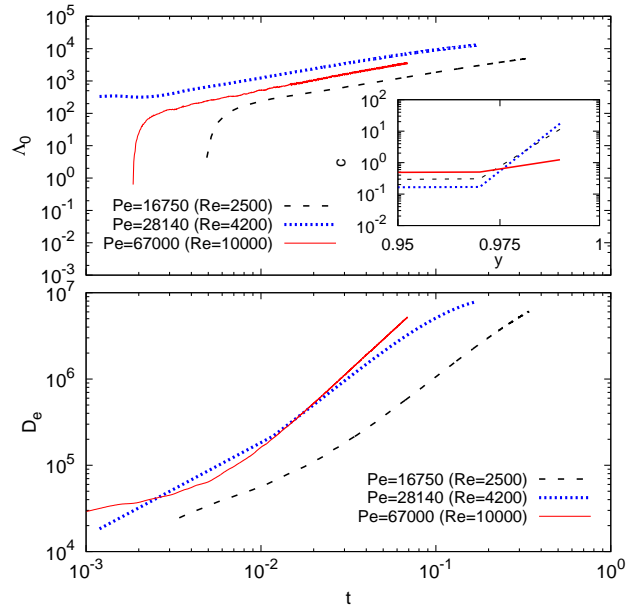}
\caption{Dispersion in upwelling turbulent flows for the same parameters as the downwelling case. Top: mean drift $\Lambda_0$ (inset: cell concentration profile close to the wall). Bottom: effective diffusivity $D_e$. Note that in the turbulent case neither of the statistical measures reaches a steady value.}\label{turb_up}
\end{figure}

The upwelling turbulent case, presented in figure \ref{turb_up}, was also investigated for the same parameter values as the downwelling case of section (3\ref{turb_down}). We see that dispersion measures do not reach steady values: both $\Lambda_0$ and $D_e$ grow monotonically with time for the duration of the simulation run. The inset to the top figure \ref{turb_up} shows the cell concentration profiles displaying a thin band of gyrotactic accumulation at the walls. Cells that end up in this band are subject to strong dispersal by the large mean shear rate at the wall. Diffusion, whether due to turbulence or swimming, is not strong enough to balance the shear-induced migration towards the wall, as is the case for laminar flow, leading to non-steady dispersion over the course of the simulations.

However, the upwelling dispersion results presented above are not realistic for swimming species that are negatively buoyant; accumulations of negatively buoyant cells at the walls can result in instability and the formation of downwelling flow and descending plumes near the walls \cite{Kessler85}. These instabilities will differ in their extent in the laminar and turbulent upwelling cases, but we expect buoyancy driven wall flows to ensue after accumulation in both regimes.

\subsubsection{The effects of cell elongation}

So far we have assumed that the gyrotactic cells are spherical, setting the eccentricity parameter $\alpha_0$ to zero. Here we present simulations obtained with a non-zero eccentricity, $\alpha_0=0.8$, for the cases $($Pe,\,$\beta)=(670,\,6.7)$ (laminar) and $(28140,\,55)$ (turbulent).
\begin{figure}[h!]
\centering
\includegraphics[width=84mm]{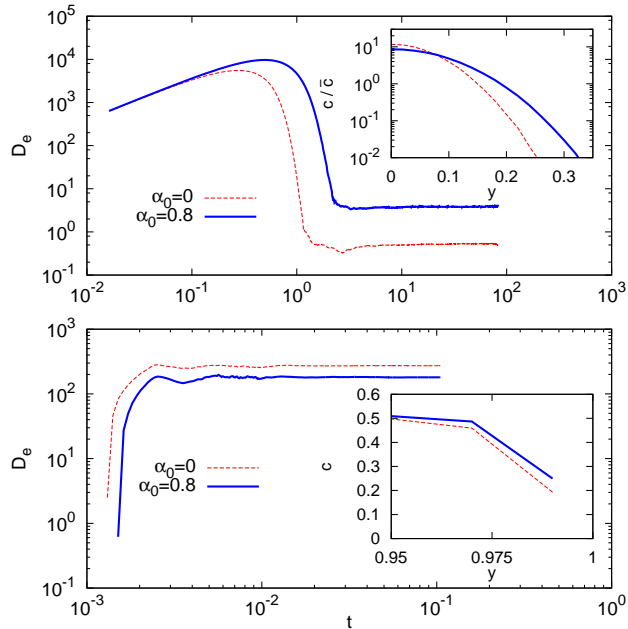}
\caption{The effect of cell elongation on dispersion. DNS simulations of the effective diffusivity $D_e$ performed for spherical ($\alpha_0=0$) and elongated ($\alpha_0=0.8$) cells for downwelling flows. Top: laminar flow, with Pe$=670$ and $\beta=6.7$; bottom: turbulent flow, with Pe$=28140$ and $\beta=55$. Even for the large $\alpha_0$ value considered, qualitative effects are not observed and significant quantitative differences are apparent only in the laminar case (see text and supplementary materials).}\label{turb_ell}
\end{figure}
We chose the relatively large value of $\alpha_0=0.8$ to emphasise the effect of eccentricity. Results for the effective diffusivity for downwelling flow are shown in figure \ref{turb_ell} for the laminar and turbulent cases. The insets in the figure display the concentration profiles. We see that in the laminar case the effect of eccentricity is to broaden the profile, as observed in Bearon {\it et al.} \citep{BearonHazelThorn11}. This broader distribution results in an increased value of effective diffusivity (cells sample more of the shear profile). In the turbulent case of figure \ref{turb_ell} (bottom) there is a much smaller broadening effect.  The effect of cell elongation on other statistical measures for the parameters considered here is marginal, see supplementary materials. If realistic values of biflagellate eccentricity ($\alpha_0=0$-$0.3$ \citep{OMalleyBees11}) are used, predictions for the dispersion of biased swimmers are not very different from those obtained here for spherical cells.

\subsection{Long-time dispersion of gyrotactic cells as a function of $Pe$ ($Re$) across the laminar-turbulent transition \label{longdisp}}

Having analysed the full time-dependence of gyrotactic cell dispersion, we concentrate here on its long-time behaviour. In this limit, laminar DNS results can be compared with predictions from analytical theory for drift $\Lambda_0$ and diffusivity $D_e$ as functions of Pe (for given $\beta$). The theory requires as inputs expressions for the mean swimming direction, ${\bf q}$, and diffusivity tensor, ${\bf D}$, obtained from microscopic stochastic models. We test two alternative microscopic models: Fokker-Planck (FP) and generalised Taylor dispersion (GTD). The models predict qualitatively different functional forms for the components of ${\bf D}$ as a function of the dimensionless shear $\sigma$. Correspondingly, predictions for $\Lambda_0($Pe$)$ and $D_e($Pe$)$ obtained with the two models also differ.
\begin{figure}[h!]
\centering
\includegraphics[width=110mm]{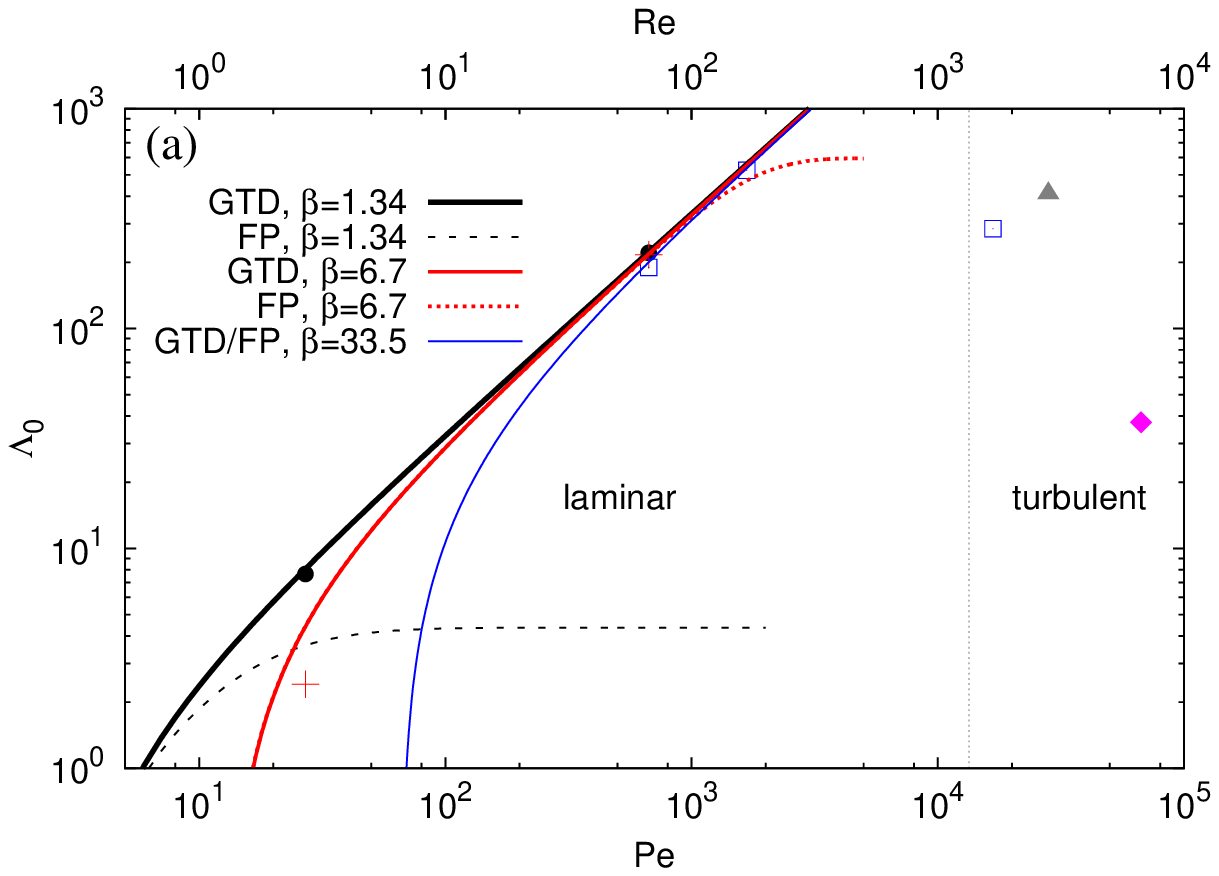}\\
\includegraphics[width=110mm]{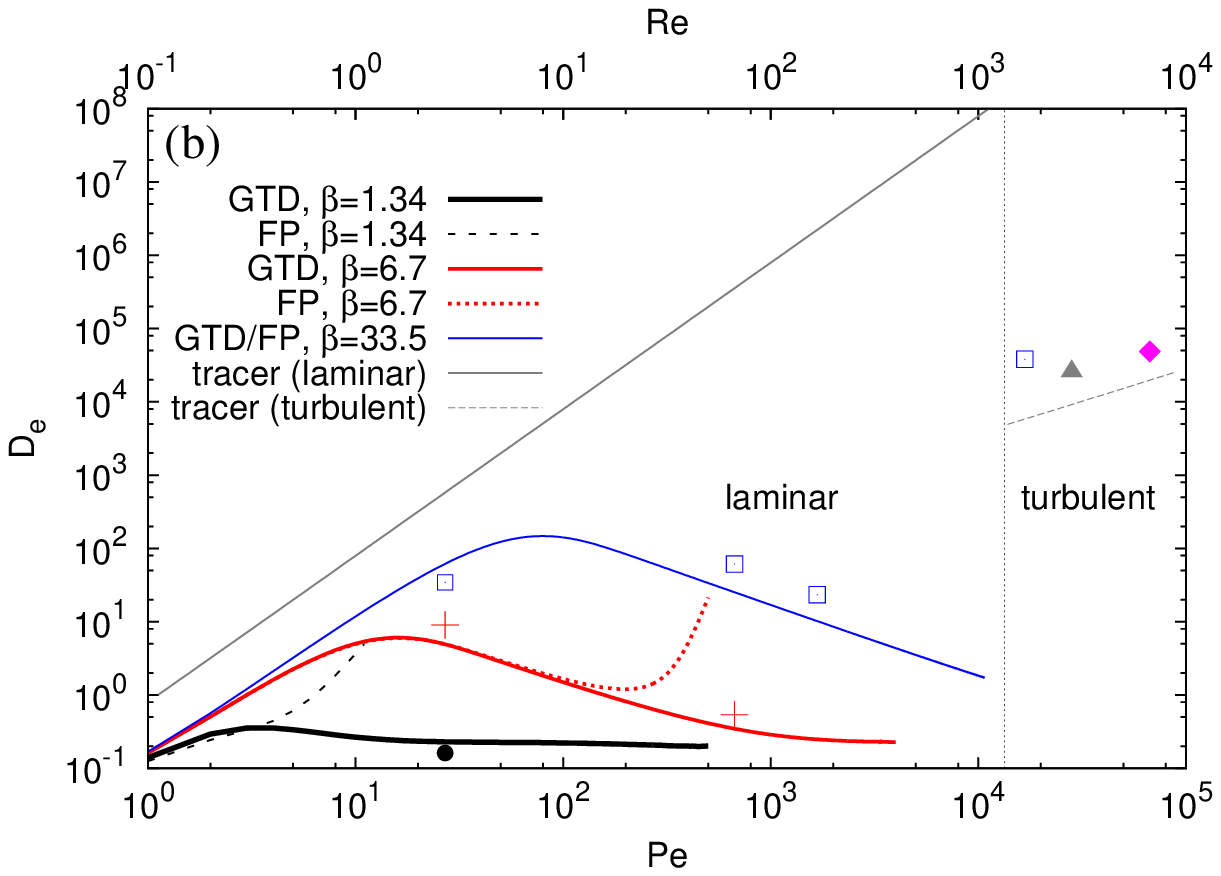}
\caption{DNS results for long-time (a) drift, $\Lambda_0$, and (b) diffusivity, $D_e$, as a function of Pe for gyrotactic cells in downwelling flows. Values of $\beta$ are: $1.34$ ($\CIRCLE$), $6.7$ ({\color{red}$+$}), $33.5$ ({\color{blue}$\square$}), $55$ ({\color{grey}$\blacktriangle$}), $67$ ({\color{fuchsia}$\vardiamond$}). The DNS results are compared with predictions from the Bees \& Croze dispersion theory, using the FP and GTD microscopic models, plotted as lines for the same $\beta$ values as the DNS, see legend. Simulation results compare well with GTD predictions for all $\beta$, but,  for Pe$\gtrsim1$, they are incompatible with FP predictions for small $\beta$. Also plotted in (b) are the classical results for passive tracers. In laminar flows, $\Lambda_0=0$ and $D_e\sim$Pe$^2$, in sharp contrast with our predictions for swimmers. In turbulent flow, we still have $\Lambda_0>0$ for swimmers, though it appears to decay to zero with increasing Pe. The turbulent $D_e$ for swimmers is close to the passive case; see text.}\label{disp_lamturb}
\end{figure}

The predictions from the analytical theory are shown as curves in figure \ref{disp_lamturb}a, b, for fixed values of $\beta$; the corresponding DNS results for selected Pe and the same values of $\beta$ are shown as points. It is clear that, for the GTD predictions, the agreement between theory and simulation for $\Lambda_0$ is very good. The GTD prediction is that, for fixed $\beta$, $\Lambda_0$ increases with Pe, tending to a linear behaviour for large enough Pe. This is in contrast to passive tracers, for which $\Lambda_0=0$.  For small Pe the FP prediction coincides with that of GTD, but, for larger Pe, $\Lambda_0$ tends to a constant instead of growing with Pe. The smaller the value of $\beta$, the greater is the difference between the predictions of the two microscopic models (as for concentration profiles). A similar trend is seen for the effective diffusivity $D_e$, shown as a function of Pe in \ref{disp_lamturb}b. We see that for $\beta=33.5$, both GTD and FP predict a diffusivity rising and then falling with Pe, consistent with DNS results. For smaller values of $\beta$, however, the FP and GTD predictions are different: GTD predicts a decrease of $D_e$ at large Pe, while FP predicts a (second) rise. The DNS results are not compatible with this rise, but agree well with the GTD prediction. We remark that this dispersion behaviour is unique to swimming cells. It is due to gyrotactic bias and the ensuing accumulations that change the distribution of cells and thus their dispersion, as discussed above.  Also shown in figure \ref{disp_lamturb}b is the diffusivity for passive tracers, which grows as $D_e\sim$ Pe$^2$ \cite{Taylor53},
without the decrease at large Pe we predict for gyrotactic swimmers.

In earlier sections we have seen how the effect of gyrotaxis on accumulation and dispersion is more subtle in turbulent flows. DNS results for long-time dispersion reflect these changes and are also summarised in figure \ref{disp_lamturb}.  An analytical theory of turbulent swimming dispersion has yet to be formulated, so we compare only with theory for passive tracers.  It is seen for all simulations with $\beta=33.5$, that the value of $\Lambda_0$, which was growing with Pe in the laminar regime, drops a little just beyond the transition to turbulence. On the other hand, the diffusivity $D_e$, which was falling at large laminar Pe, suddenly acquires a sizeable value. The turbulent shear profile and fluctuations alter the distribution of cells. The effect of gyrotaxis is thus weakened but still measurable in the dispersion, which is similar but not identical to that of passive tracers. The fractional drift above the mean flow speed $U=(2/3) U_c$, is given by $\delta\equiv \lim_{t\to\infty} (1/U)[d m_1/dt-U]=(3/2)\Lambda_0/\rm{Pe}$; this is rather small for large turbulent Pe values. Nevertheless, it is remarkable that drift should not be zero in a turbulent flow, as it is for passive tracers. For very large values of Pe, we expect gyrotaxis to have practically no effect on dispersion: the time-averaged concentration profile will be well-approximated as uniform and $\Lambda_0=0$. Indeed our results suggest that $\Lambda_0\to0$ for increasing Pe in the turbulent regime.

\section{Discussion \label{Discussion}}

We have studied the dispersion of gyrotactic swimming algae in channel flows by direct numerical simulations (DNS) and analytical theory. This is the first study to evaluate cell distributions and statistical measures of gyrotactic cell dispersion (drift above mean flow $\Lambda_0$, effective diffusivity $D_e$ and skewness $\gamma$) for flows on either side of the laminar-turbulence transition. We find unique cell accumulations and dispersion with non-zero drift and a non-monotonic diffusivity with Peclet number, Pe, with qualitative differences for upwelling and downwelling flows. The dispersion behaviour is remarkable and unique to gyrotactic swimming cells; passive tracers are transported at the mean flow rate ($\Lambda_0=0$), the diffusivity increases with Pe ($D_e\sim$Pe$^2$) \citep{Taylor53}, and dispersion is insensitive to channel orientation. In the laminar downwelling regime, simulation results were compared with the predictions derived from a recently formulated general analytical theory of swimming dispersion \citep{BeesCroze10, BearonBeesCroze12}, applied here for the first time to channel geometry. The theory requires as inputs expressions for cell response to the local shear in a flow, determined from microscopic models.

We have evaluated predictions based on two alternative models: the Fokker-Planck (FP) \citep{PedleyKessler90, Beesetal98} and generalised Taylor dispersion (GTD) \citep{HillBees02, ManelaFrankel03, BearonBeesCroze12} approaches. Which model is more realistic has long been a matter of debate \citep{HillBees02}. Here we find that the DNS are in excellent agreement with analytical predictions based on GTD, providing the first evidence that it is much better than the FP model at describing swimmers in flows. As well as validating analytical predictions for laminar flow, the DNS allow us to explore the industrially relevant dispersion of gyrotactic algae in turbulent channel flows. In the turbulent regime the effects of gyrtoaxis (accumulations resulting in non-zero drift) persist, but are much more subtle. Effective diffusivity in downwelling turbulent flows is similar to that of passive tracers. These are the first full direct numerical simulations of biased swimmers in turbulent channel flows; previous studies having concentrated on vortical flow \citep{DurhamClimentStocker11} or synthetic approximations of homogeneous isotropic turbulence \citep{Lewis03, ThornBearon10}.

The fact that swimming algae in channel flows distribute very differently to passive tracers has important practical consequences for the culture of swimming species. The most dramatic implications of our findings are for photobioreactors that operate using laminar flows. For example, in draft tube air-lift bioreactors, bubbles up a central draft tube (riser) mix and aerate cells, which then circulate down the channel formed between the draft tube and the surface of the reactor (downcomer; see figure \ref{bioreactors}c). The Bees \& Croze \citep{BeesCroze10} analytical theory, confirmed by simulations, predicts that gyrotactic swimming algae will be focused more and more sharply at the center of the downcomer as Pe (the flow rate for fixed channel width, or Re) is increased. For example, considering a flow with Pe$=1675$ (Re$=250$) and the swimming Peclet number $\beta=33.5$ (H$=5$cm), we predict that the concentration at the walls is a vanishingly small fraction of the mean, given by $c/\bar{c}\approx \exp{[-\rm{Pe} \lambda/(2 \beta)]}\sim10^{-22}$ (using $\lambda=2.2$ for {\it Chlamydomonas augustae}; see approximations in Appendix A). Non-swimming cells and molecular solutes, on the other hand, would be uniformly distributed across the tube, $c/\bar{c}=1$. As we can reasonably assume that the probability of cell adhesion to the walls is proportional to the concentration there, we predict that surface fouling by gyrotactic cells will be markedly reduced relative to non-swimming cells. Fouling can be a big problem in closed bioreactors because cells buildup can prevent light penetration and thus growth, and in extreme cases can clog reactor conduits
\citep{BrennanOwende10}.

The peculiar dispersion of biased swimmers will also affect growth in bioreactors. Our results indicate that gyrotactic swimmers in a downwelling flow drift faster and diffuse less than passive tracers, which travel at the mean flow. We can make experimentally measurable predictions for the dispersion of {\it Chlamydomonas augustae}, the alga whose swimming parameters (reorientation time $B$, rotational diffusivity $d_r$, swimming speed $V_s$) we have used in this study. For example, in the case with $\beta=6.7$ and Pe$=670$ (e.g. a realistic, small air-lift with $H=1$ cm and $U_c=1$ cm/s) the fractional drift above the mean flow, $\delta=(3/2)\Lambda_0/\rm{Pe}$, is estimated to be $\delta=0.48$. In other words {\it C. augustae} cells drift about $50\%$ faster than passive tracers, such as nutrients. For the nondimensional effective diffusivity we predict $D_e=0.8$: the axial diffusivity of cells is smaller than the cell diffusivity scale ($V_s^2/d_r=1.49\times10^{-3}$ cm$^2$/s) in the absence of flow. To compare these predictions with those for passive molecular solutes, we consider the dispersion of CO$_2$ (molecular diffusivity $D_t=1.6\times10^{-5}$ cm$^2$/s), a vital nutrient for algae. In the flow under consideration, cells have Pe$=670$, but CO$_2$, with its smaller diffusivity, has Pe$_t=$Pe$\, V_s^2/(d_r D_t)=6.2\times10^{4}$. Thus, the non-dimensional tracer diffusivity is $D_{e,t}\approx(8/945)$Pe$_t^2\sim 10^7$, from the Taylor-Aris result. Re-dimensionalising swimming and solute effective diffusivities by multiplying by diffusivity scales, we find that the axial diffusivity of CO$_2$ along a channel is $\approx10^5$ times greater than that of {\it C. augustae}. This dramatic differential dispersion of cells and nutrients could have important consequences for swimming cell growth in reactors.

In turbulent downwelling flows, which may also be realised in air-lifts, the effects gyrotactic swimming are much less pronounced. Gyrotactic depletion is not as efficient as in laminar flows: the concentration at the walls is at best half of the mean concentration. We thus expect significant fouling in air-lifts under turbulent regimes. The influence of gyrotaxis on dispersion is also weaker than in the laminar case. For {\it C. augustae} in a turbulent channel flow with $\beta=55$ ($H=8.4$cm) and Pe$=28140$ (Re$=4200$, $U_c=5$cm/s), the DNS predict a fractional drift of $\delta=0.015$. This is very small, and may be neglected for short channels. As shown in figure \ref{disp_lamturb}b, the effective diffusivity is of the same order as that of a passive tracer such as CO$_2$.


The excellent agreement between DNS and predictions for dispersion obtained using generalised Taylor dispersion theory, provides a first confirmation that the swimming dispersion theory of Bees \& Croze \citep{BeesCroze10} is robust. If they are to be useful in bioreactor engineering design, however, the theoretical and numerical predictions for dispersion need to be tested against experiments. Work is in progress to test GTD predictions for pipe flow with the alga {\it Dunaliella} \citep{CrozeBearonBees12}. It would also be interesting to test experimentally the numerical and analytical predictions for channel flow presented herein, exploring in particular the laminar-turbulent transition region. Since tubes are often arranged horizontally in bioreactor designs, it will be interesting to study the effect of tube orientation on biased swimmer dispersion. Croze {\it et al.} \citep{CrozeAshrafBees10} have carried out experiments on the dispersion of {\it Chlamydomonas augustae} in horizontal pipe flow for low Pe. They found a complex transport mediated by sheared bioconvection patterns and suggested that such cell-driven flows could alter the transition to turbulence. It would be interesting to study this transition experimentally and with the aid of simulations such as presented here, including the effects of cell buoyancy (especially for small Pe flows) \cite{BearonBeesCroze12}.

More generally, our study could be extended to open channel flows, such as those present in pond bioreactors, channels, waterfalls and rivers. The swimming dispersion effects we have explored must also exist whenever biased swimmers are subject to shear in the ocean and lakes. Significantly, about $90\%$ of species implicated in the formation of harmful algal blooms swim using flagella \citep{Smayda97}. A recent study has shown that {\it Heterosigma akashiwo} is gyrotactic and could be responsible for thin layer formation \citep{Durhametal09}.  However, the role of biased swimming in the dispersion and ecology of algae remains largely unexplored.

\section*{Acknowledgements}

We thank Rachel Bearon for comments and discussions, and Miriam La Vecchia for help with implementing the formulation based on quaternions.  Discussions with Victoria Adesanya, Stuart Scott and Alison Smith on photobioreactors are also acknowledged. Computer time provided by SNIC (Swedish National Infrastructure for Computing) is gratefully acknowledged. OAC and MAB are thankful for support from the Carnegie Trust and EPSRC (EP/D073398/1 and EP/J004847/1).

\appendix{Microscopic model solutions \label{appA}}

\subsection{Numerical solutions for the Fokker-Planck and generalised Taylor dispersion models}

${\bf q}(\sigma)$ and ${\bf D}_m(\sigma)$ predicted by the FP and GTD microscopic models have the following structure:
\begin{equation}
\mathbf{q}=
\left[
\begin{array}{c}
q^x(\sigma)\\
q^y(\sigma)\\
0
\end{array}
\right];
\mbox{~~~~~~~~~~}
\mathbf{D}_m=
\left[
\begin{array}{ccc}
D^{xx}_m(\sigma)&D^{xy}_m(\sigma)&0\\
D^{xy}_m(\sigma)&D^{yy}_m(\sigma)&0\\
0&0&D^{xx}_m(\sigma)
\end{array}
\right],
\end{equation}
where we recall $m=$F represents solutions of the FP model and $m=$G those of the GTD model. Such solutions can be obtained using approximations via a Galerkin method \cite{BearonBeesCroze12}.  It is convenient to fit such solutions with the following expressions: $q^x(\sigma)= -P(\sigma; {\bf a}^x,{\bf b}^x)$; $q^y(\sigma)=-\sigma P(\sigma; {\bf a}^y,{\bf b}^y)$; $D^{xx}_m(\sigma)= P(\sigma; {\bf a}^{xx},{\bf b}^{xx})$; $D^{yy}_m(\sigma)= P(\sigma; {\bf a}^{yy},{\bf b}^{yy})$;$D^{xy}_m(\sigma)= -\sigma P(\sigma; {\bf a}^{xy},{\bf b}^{xy})$. Here  the rational function $P(\sigma; {\bf a},{\bf b})$ is given by
 \begin{eqnarray}
\label{eq:P_rat_func}
P(\sigma; \mathbf{a},\mathbf{b})=\frac{a_0 +a_2 \sigma ^2+a_4\sigma^4}{1+b_2 \sigma^2+b_4 \sigma^4},
\end{eqnarray}
 and, for $\lambda=1/(2 d_r B)=2.2$, the coefficients ${\bf a}$ and ${\bf b}$ are provided in table (\ref{abcoeff}) below. \\

\begin{table}[h!]
\begin{center}
\begin{tabular}{l||c|c|c|c|c}
 & $a_{0}$ &$a_{2}$ & $a_{4}$&$b_{2}$ & $b_{4}$\\
\hline
$\mathbf{a}^x$		&$5.7 \times 10^{-1}$		&$3.66 \times 10^{-2}$	&$0$			&$1.75 \times 10 ^{-1}$	&$1.25 \times 10^{-2}$\\
$\mathbf{a}^y$		& $2.05 \times 10^{-1}$ 	&$1.86 \times 10^{-2}$	&$0$			&$1.74  \times10 ^{-1}$	&$1.27 \times 10^{-2}$\\
$ \mathbf{a}^{xx}_G$	&$5.00 \times 10^{-2}$	&$1.11 \times 10^{-1}$	&$3.71 \times 10^{-5}$&$1.01 \times  10 ^{-1}$	&$ 1.86 \times 10^{-2}$\\
$ \mathbf{a}^{yy}_G$	&$9.30 \times 10^{-2}$	& $1.11 \times 10^{-4}$	&$0$			&$1.19\times 10 ^{-1}$	&$1.63 \times 10^{-4}$\\
$ \mathbf{a}^{xy}_G$	&$9.17\times 10^{-2}$	&$1.56\times 10^{-4}$	&$0$			&$2.81 \times  10 ^{-1}$	&$2.62\times 10^{-2}	 $\\
$ \mathbf{a}^{xx}_{F}$	&$5.60 \times 10^{-2}$	& $3.23 \times 10^{-2}$	&$1.70 \times 10^{-5}$&$2.70\times10 ^{-1}$		&$1.42 \times 10^{-4}$\\
$ \mathbf{a}^{yy}_{F}$	&$9.30 \times 10^{-2}$	& $5.73\times 10^{-4}$	&$1.85\times 10^{-3}$	&$4.96\times 10 ^{-2}$	&$1.54 \times 10^{-2}$\\
$ \mathbf{a}^{xy}_{F}$	&$1.58 \times 10^{-2}$	&$0$				&$0$			&$9.61\times 10^{-2}$		&$7.88 \times 10^{-2}$
\end{tabular}
\end{center}
\caption{Parameters used in the functional fits of solutions to the FP and GTD models used for our analytical theory predictions (from \citep{BearonBeesCroze12}).} \label{abcoeff}
\end{table}

\subsection{Approximate GTD profiles and width scaling}

Using asymptotic results derived in \citep{BearonBeesCroze12}, we derive an approximation for the concentration profiles predicted by the GTD model. These profiles are given by equation (\ref{conceq}) as $c(y)/\bar{c}= \exp{\left( \beta \int^y_0 (q^y/D_m^{yy})ds \right)}$, where $m=G$ for GTD predictions. For $\sigma\ll1$ at leading order the GTD prediction asymptotes to $q^y(\sigma)=-\sigma J_1/\lambda$, where $J_1$ is a known constant for $\lambda=2.2$. In the same limit, $D^{yy}_G(\sigma)=J_1/\lambda^2$, so that $(q^y/D^{yy}_G)_{\sigma\ll1}=-\sigma \lambda$. For $\sigma\gg1$ at leading order, $q^y(\sigma)=-(2/3)\lambda/\sigma$ and $D^{yy}_m(\sigma)=d_1/\sigma^2$, where $d_1=0.68$ for $\lambda=2.2$. Thus $q^y/D^{yy}_G\approx-\sigma \lambda$ is a reasonable approximation for all $\sigma$. Recalling for channel flow $\sigma=({\rm{Pe}}/\beta^2)y$, we see that $\beta q^y/D^{yy}_G\approx  \lambda ({\rm{Pe}}/\beta) y$. Inserting this expression in the equation above and integrating gives the Gaussian profile
\be
c(y)\approx \exp{(y/y_0)^2},
\ee
where
\be
y_0\approx [(2/\lambda)(\beta/\rm{Pe})]^{0.5}
\ee
is the width of the profile. This scaling is observed in the concentration profiles obtained from DNS; see main text.

\section*{Supplementary materials}

\subsection*{S1. Numerical methods for the simulations}

The numerical code is an efficient pseudo-spectral solver for the three-dimensional incompressible Navier-Stokes equations (equations 2.3 or 2.6 in the main text) with a particle tracking algorithm \cite{Sardinaetal12} adapted to swimming micro-organisms. The velocity components of the fluid phase are expanded in both $x$ (streamwise) and $z$ (spanwise) direction with Fourier modes, whereas Chebyshev polynomials are employed in the wall-normal $y$-direction. To advance the Navier-Stokes equations in time, we use a fourth order Runge-Kutta algorithm. Periodic boundary conditions are assumed in $x$ and $z$, with no-slip at both walls $y=\pm 1$. More details about the code are given in \cite{Chevalieretal07}. For the DNS with laminar flow, the flow is not simulated, but is taken to be the analytic parabolic Poiseuille solution to the Navier-Stokes equations .

The micro-organisms are treated as point particles evolved by means of a Lagrangian solver. The fluid velocities and the components of the velocity gradient tensor are interpolated from the Eulerian grid onto the particle positions using a tri-linear interpolation. The time advancement of the particle uses the same Runge-Kutta algorithm as the time-advancement of the fluid except for the noise which is integrated by an Euler-Marayuma method \cite{Higham01}. Equation (2.5) of the main text is solved using quaternions rather than angles \cite{LundellCarlsson10}. We perform numerical simulations of turbulent plane Poiseuille flow at constant mass flux. The Reynolds number, domain size and resolution used for the results presented here are reported in table~\ref{table:turb}. Note that resolution and box size are chosen following the classic results for turbulent channel flow in \cite{Kimetal87, Moseretal99}.

\begin{table}
\centering
{\begin{tabular}{ c| c c c c }
Box & $Re$ & $Re_{\tau}$  & $L_x \times L_y \times L_z$ &   $N_x \times N_y \times N_z$ \\
\hline
A & 2500 & 120 & $4\pi\times 2 \times 2\pi$ & $64 \times 65 \times 64$ \\
B & 4200 & 180 & $4\pi\times 2 \times 2\pi$ & $128 \times 129 \times 128$ \\
C& 10000 & 390 & $2\pi \times 2 \times 4\pi/3 $ & $ 512 \times 193  \times 256$ \\
\end{tabular}}
\caption{Large-scale and turbulent (or friction) Reynolds number for the simulations presented here. The table also reports the simulation box size in units of $H$, the channel half width, and the resolution used for the simulations.}
\label{table:turb}
\end{table}

In the table, we report both the large-scale Reynolds number defined by $Re = U_c H/\nu$ where
$U_c$ is the centerline streamwise velocity for the laminar flow of same mass flux, and the turbulent Reynolds number $Re_{\tau} = u_{\tau}  H/\nu$. The latter is defined by the friction velocity $u_{\tau} = \sqrt{\sigma_{\rm w}}$ and
the half-channel width, where $\sigma_{\rm w} \equiv \nu [\partial U/\partial y]_{\rm wall}$ is the shear stress at the wall \cite{Pope00}.

Note that the simulation domain is shorter than the distance travelled by the swimmers during dispersion; also, for large times, the area occupied by the cells becomes longer than our computational box. However, the computational domain can be considered long enough for the velocity correlation to be negligible at a distance of the order of the box length $L_x$. Therefore, a very long domain for the swimmers can be assembled by means of copies of a single Eulerian
computational box of length $L_x$ by using the method of repeated domains described in \cite{Picanoetal09}.

\subsection*{S2. Evalutation of the analytical dispersion measures}

To evaluate the integrals in (2.20) and (2.22), it was simpler to numerically evaluate an equivalent system of steady differential equations (e.g. use the numerical solution to $d Y_0^0/dy = \beta (q^y/D^{yy}) Y_0^0$, instead of (2.21)). The system of ODEs was solved using the Matlab {\it bvp4c} routine. Solutions to the system were found for using both the Fokker-Planck and GTD expressions given in Appendix A. For comparison with the analytical predictions, concentration profiles from DNS were normalised by dividing by the area under the profile (the number of cells), evaluated using the Matlab {\it trapz} routine. 
\clearpage

\subsection*{S3. Supplementary results}

\begin{figure}[!h]
\centering
(a)\includegraphics[width=0.6\linewidth]{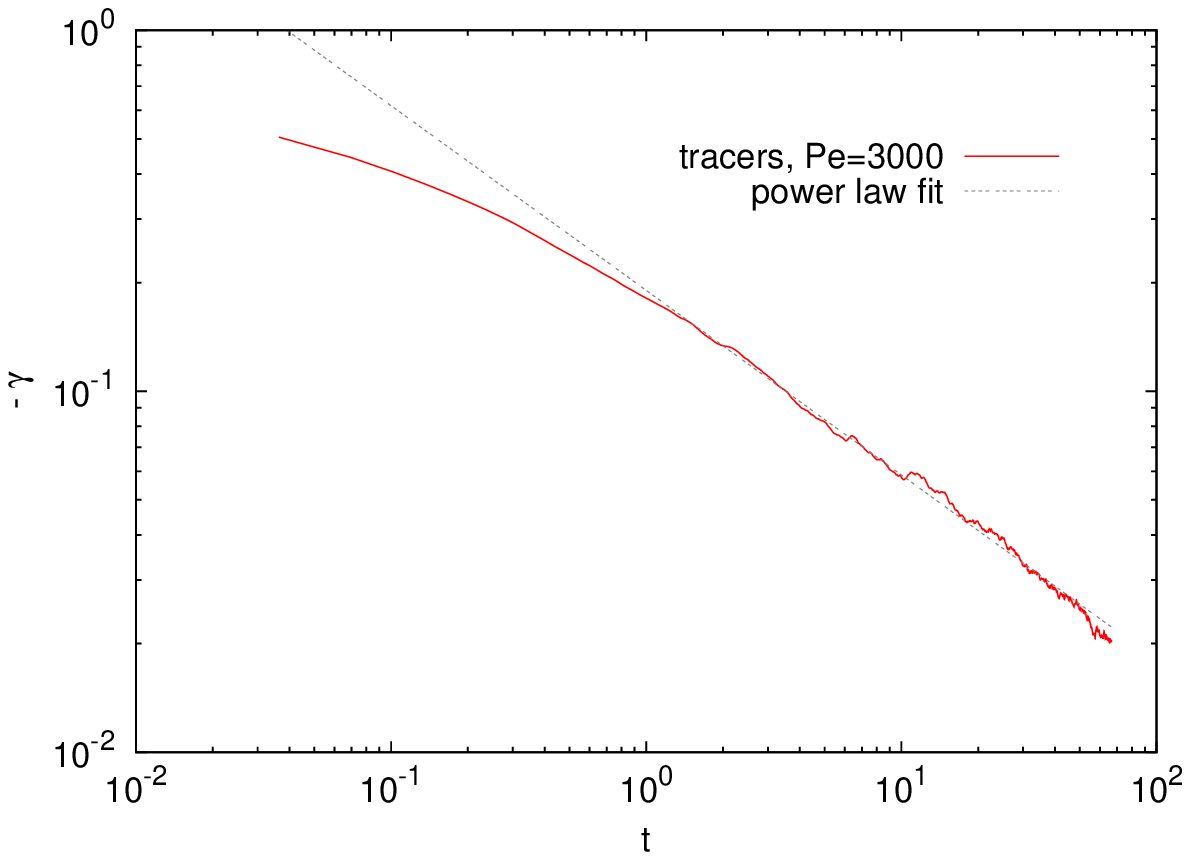}
(b)\includegraphics[width=0.6\linewidth]{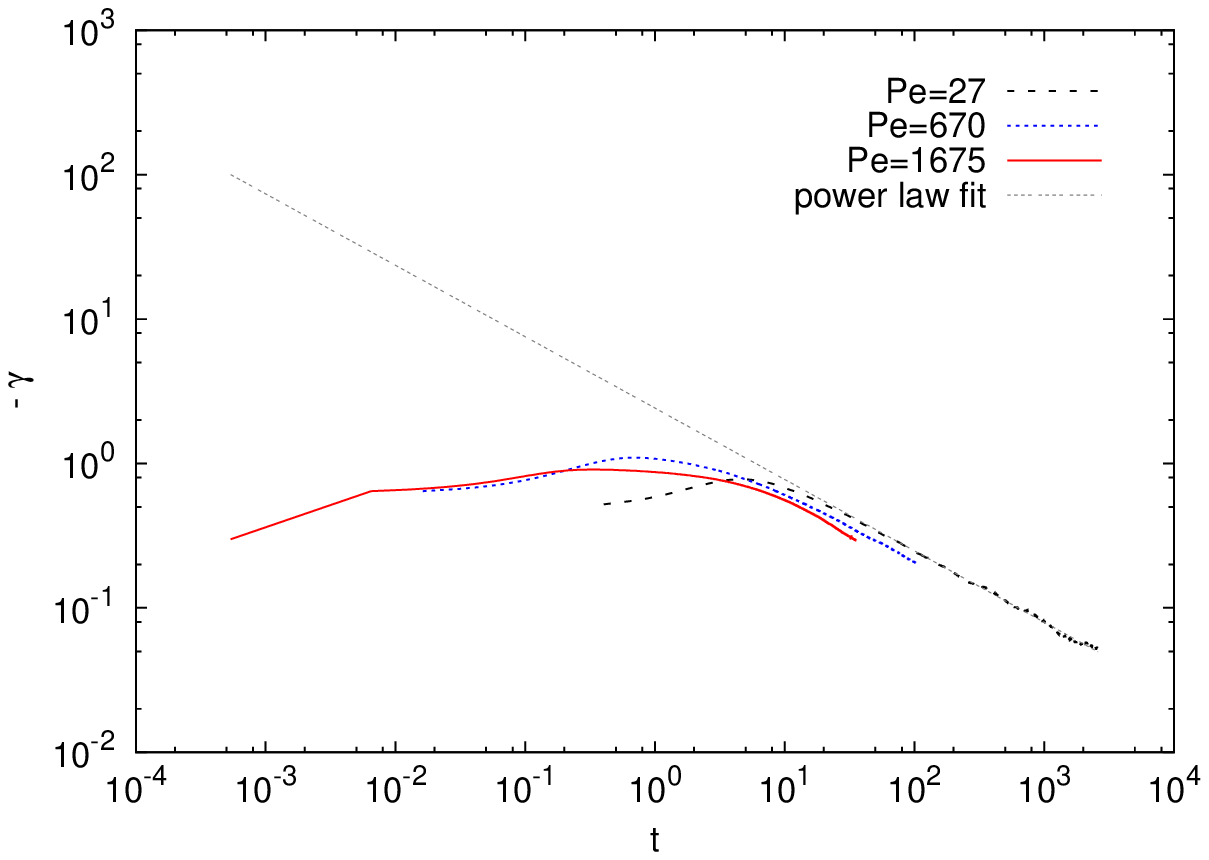}
\caption{Approach to zero of $-\gamma$ (the negative of the skewness so we can use a log-log plot) for (a) passive tracers and (b) gyrotactic swimmers  for the same values of Pe considered in figures 3 and 6 of the main text, respectvely. We fit the passive case and the swimmer case with Pe$=27$ with the power law $\gamma=\gamma_0 t^{-\alpha}$ where $\gamma_0$ and $\alpha$ are constants. For tracers we find $\gamma_0=0.19$ and $\alpha=0.51$ and for swimmers $\gamma_0=2.42$ and $\alpha=0.49$. The pre factors are different, but both  exponents are compatible, as predicted by Bees \& Croze \cite{BeesCroze10} who calculated that the skewness for swimmers should approach zero as $\sim t^{-0.5}$ at long times, as it does for passive tracers \cite{Aris56}.}\label{skew fit}
\end{figure}
\begin{figure}[h!]
\centering
\includegraphics[width=0.6\linewidth]{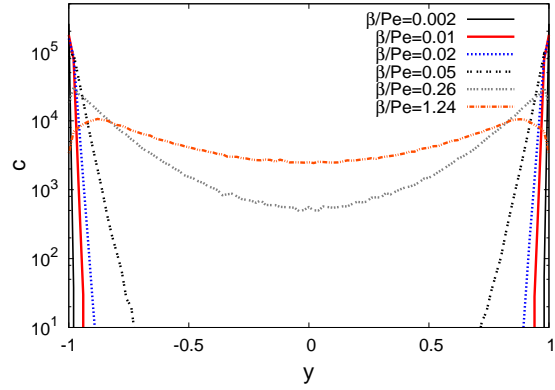}
\caption{Gyrotactic cell concentration profiles (not normalised) for upwelling laminar flows. Just as in the laminar case, gyrotactic accumulations become more pronounced as the ratio $\beta/$Pe  is decreased, but in the upwelling case these accumulations are at the walls.}\label{upwellc}
\end{figure}
\begin{figure}[h!]
\centering
(a)\,\includegraphics[width=0.4\linewidth]{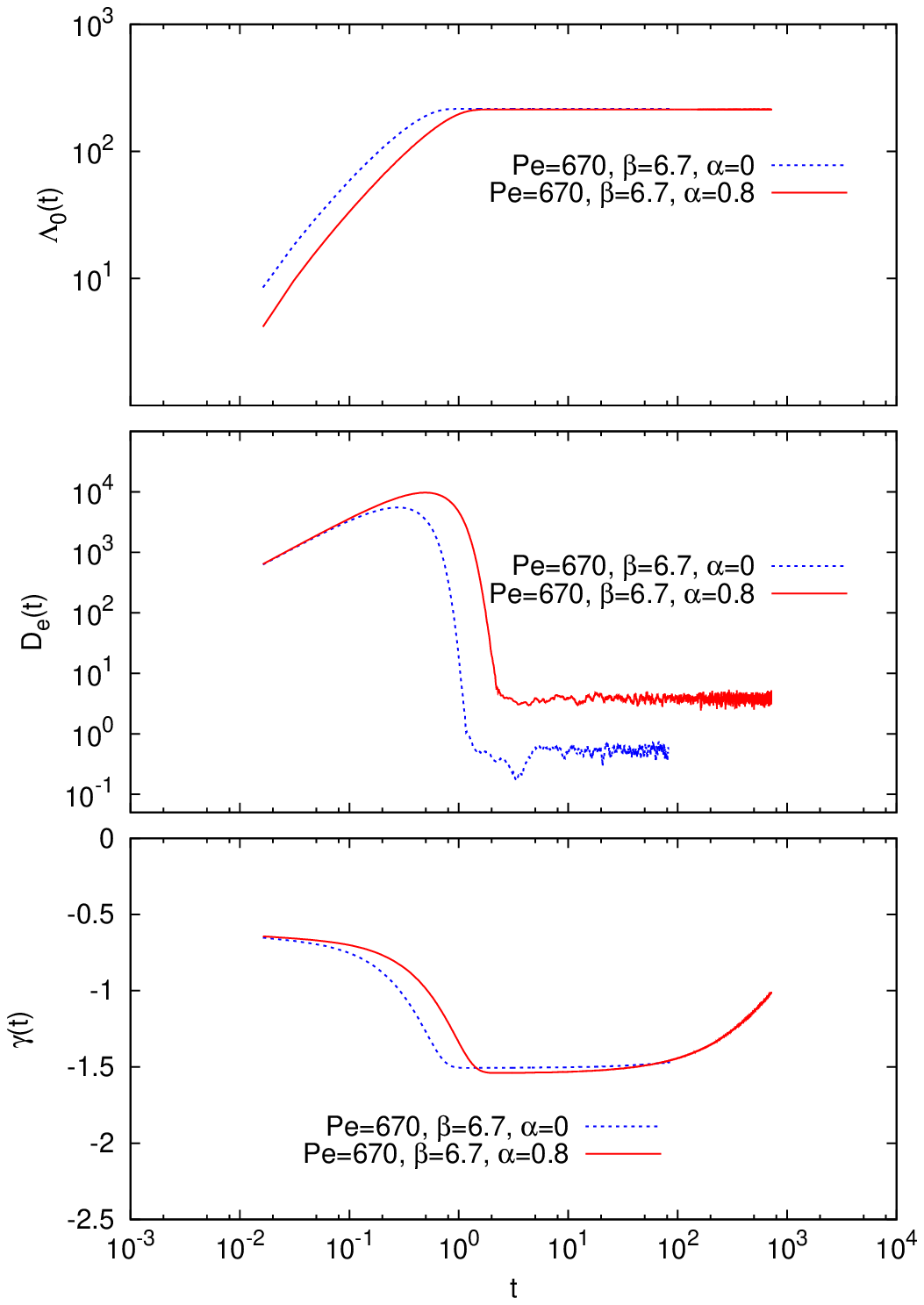}
(b)\,\includegraphics[width=0.4\linewidth]{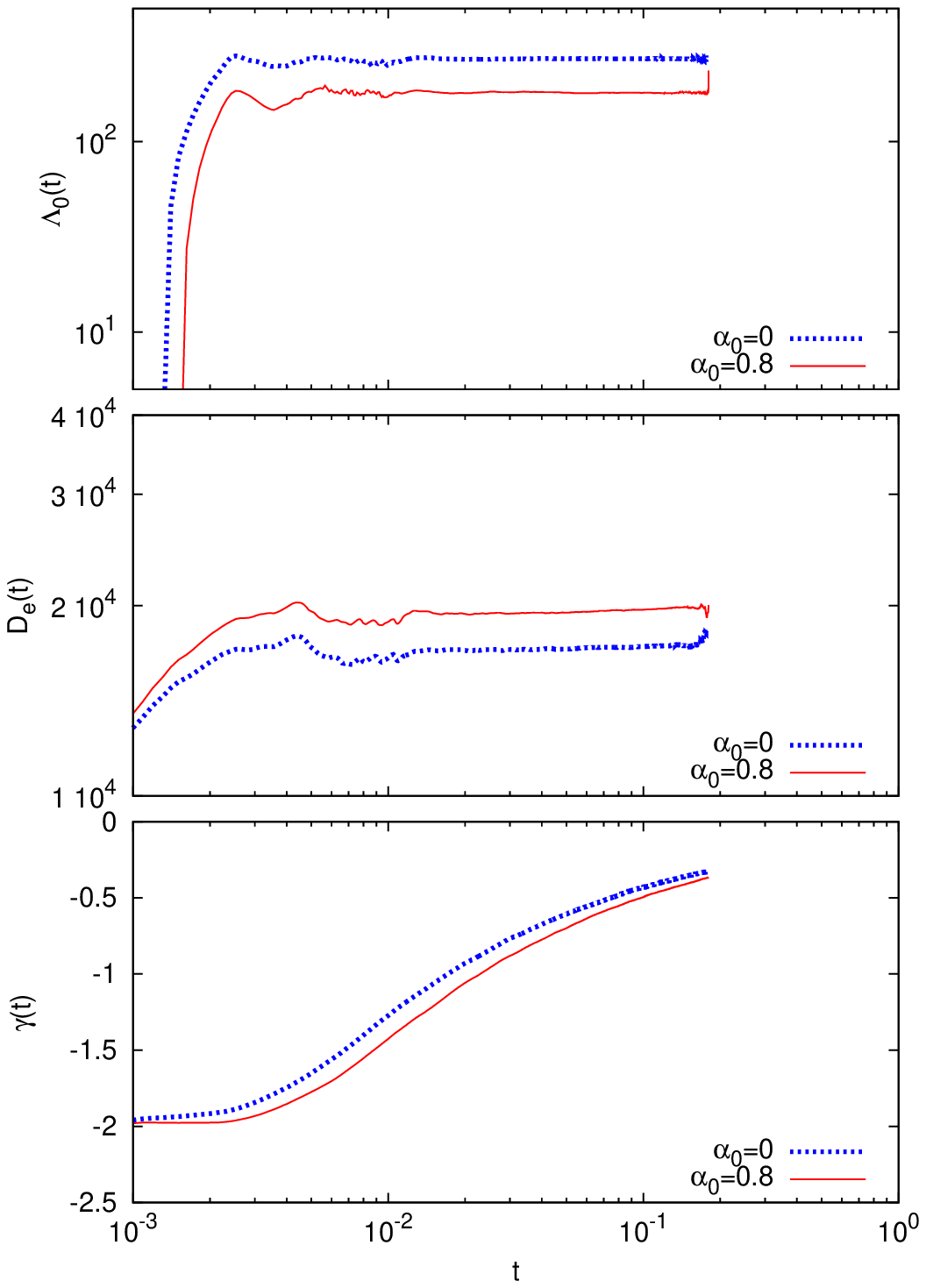}
\caption{Effect of cell elongation on dispersion measures for downwelling flows: (a) laminar (b) turbulent. Even for the unrealistically large value of eccentricity used ($\alpha_0=0.8$), the effect of elongation is small apart from the diffusivity in the laminar case, as discussed in the main text.}\label{upwellc}
\end{figure}

\newpage


\end{document}